\documentclass[structabstract]{aa}
\usepackage{graphicx}
\usepackage{txfonts}
\usepackage{natbib}
\usepackage{longtable}
\usepackage{amsmath}

\bibpunct[]{(}{)}{;}{a}{}{,}

\begin{document}

   \title{Investigation of the rotational spectrum of CH$_3$$^{17}$OH and its tentative detection 
          toward Sagittarius B2(N)\thanks{Electronic supplementary material for this work 
          can be found at https://doi.org/10.5281/zenodo.12581728}}

   \author{
           Holger S.~P. M{\"u}ller\inst{1}
           \and
           Vadim V. Ilyushin\inst{2}
           \and
           Arnaud Belloche\inst{3}
           \and
           Frank Lewen\inst{1}
           \and
           Stephan Schlemmer\inst{1}
           }

   \institute{
              Astrophysik/I.~Physikalisches Institut, Universit{\"a}t zu K{\"o}ln,
              Z{\"u}lpicher Str. 77, 50937 K{\"o}ln, Germany\\
              \email{hspm@ph1.uni-koeln.de}
              \and
              Institute of Radio Astronomy of NASU, Mystetstv 4, 61002 Kharkiv, Ukraine\\
              \email{ilyushin@rian.kharkov.ua}
              \and
              Max-Planck-Institut f\"ur Radioastronomie, Auf dem H\"ugel 69, 
              53121 Bonn, Germany
              }

   \date{Received 6 June 2024 / Accepted 28 June 2024}
 
  \abstract
{Methanol is an abundant and widespread molecule in the interstellar medium. The column 
 density of its $^{18}$O isotopolog, CH$_3$$^{18}$OH, is in some star-forming regions so high 
 that the search for CH$_3$$^{17}$OH is promising. But only very few transition frequencies 
 of CH$_3$$^{17}$OH with a microwave accuracy have been published prior to our investigation.}
{We want to extend the very limited rotational line list of CH$_3$$^{17}$OH to be able 
 to search for this isotopolog in the interstellar medium.}
{We recorded the rotational spectrum of CH$_3$$^{17}$OH between 38 and 1095~GHz employing a 
 methanol sample enriched in $^{17}$O to 20\%. A torsion-rotation Hamiltonian model based on 
 the rho-axis method was employed to fit the data, as in our previous studies. 
 We searched for rotational transitions of CH$_3$$^{17}$OH in the imaging spectral line survey 
 ReMoCA obtained with the Atacama Large Millimeter/submillimeter Array (ALMA) toward the 
 high-mass star-forming region Sgr~B2(N). The observed spectra were modeled under the assumption 
 of local thermodynamic equilibrium (LTE).}
{The assignments cover $0 \le J \le 45$, $K_a \le 16$, and mainly the $\varv_{\rm t} = 0$ and 1 
 torsional states. The Hamiltonian model describes our data well. The model was applied to derive 
 a line list for radio-astronomical observations. 
 We report a tentative detection of CH$_3$$^{17}$OH along with secure detections of the more 
 abundant isotopologs of methanol toward Sgr~B2(N2b). The derived column densities yield 
 isotopic ratios $^{12}$C/$^{13}$C = 25, $^{16}$O/$^{18}$O = 240, and $^{18}$O/$^{17}$O = 3.3, 
 which are consistent  with values found earlier for other molecules in Sgr~B2.}
{The agreement between the $^{18}$O/$^{17}$O isotopic ratio that we obtained for methanol 
 and the $^{18}$O/$^{17}$O ratios reported in the past for other molecules in Sgr B2(N) 
 strongly supports our  tentative interstellar identification of CH$_3$$^{17}$OH. 
 The accuracy of the derived line list is sufficient for further radio astronomical 
 searches for this methanol isotopolog toward other star-forming regions.}


\authorrunning{H.~S.~P. M\"uller et al.}
\titlerunning{Rotational spectroscopy of CH$_3$$^{17}$OH}

\maketitle
\hyphenation{For-schungs-ge-mein-schaft}


\section{Introduction}
\label{intro}

As one of the most abundant molecules in space, interstellar methanol was detected in the early days 
of radio astronomy \citep{det_CH3OH_1970}. It has been found in various astronomical objects, and 
several of its minor isotopic species have been detected as well, see for example the CH$_3$OH 
documentation on the Molecules in Space page\footnote{https://cdms.astro.uni-koeln.de/classic/molecules} 
of the Cologne Database for Molecular Spectroscopy \citep[CDMS,][]{CDMS_2001,CDMS_2005,CDMS_2016}. 
The $^{16}$O/$^{18}$O ratio is 500 on Earth \citep{iso-comp_2016}, $\sim$560 in the local interstellar 
medium (ISM), and $\sim$250 in the Galactic center \citep{isotopic_ratios_1994}. Nevertheless, 
the high methanol abundances in star-forming regions led to a detection of CH$_3$$^{18}$OH toward 
the Galactic molecular cloud Sagittarius (Sgr) B2 fairly early on \citep{det-CH3O-18-H_1989} with 
a reported $^{16}$O/$^{18}$O ratio of $\sim$210. In fact, lines of CH$_3$$^{18}$OH were employed 
in a line survey with the Atacama Large Millimeter/submillimeter Array (ALMA) to infer the abundance 
of CH$_3$OH toward the low-mass protostar IRAS 16293$-$2422 because of the large opacity of CH$_3$OH 
and $^{13}$CH$_3$OH lines \citep{PILS_2016}. Some of us carried out a study of alkanols and alkanethiols 
\citep{ROH_RSH_2016} toward Sgr~B2(N2). The emission lines of CH$_3$$^{18}$OH were sufficiently strong 
that the identification of CH$_3$$^{17}$OH lines looked promising. After all, the $^{18}$O/$^{17}$O 
ratio on Earth is as low as 5.5 \citep{iso-comp_2016}, $\sim$4.1 in the solar neighborhood 
\citep{18O-17O_nearby_2005}, and only $\sim$3.0 in the Galactic center \citep{18O-17O_Gal_ratio_2020}. 
But it was not possible to draw any conclusion because the CH$_3$$^{17}$OH rest frequencies were 
very uncertain as a result of a very limited amount of available laboratory data with microwave 
accuracy. To the best of our knowledge, there is only one such publication with a very limited 
set of rotational transition frequencies of CH$_3$$^{17}$OH by \citet{CH3O-17-H_rot_1991}. 
Somewhat later, \citet{CH3O-17-H_FIR_2010} carried out an investigation of the rotational and 
torsion-rotation spectrum of CH$_3$$^{17}$OH in the far-infrared region, which was extended 
shortly thereafter into the mid-infrared region, where higher torsional bands and the 
CO-stretching band were investigated \citep{CH3O-17-H_FIR_MIR_2011}. The assigned transition 
frequencies of both studies were provided as supplementary material to the later publication.

We have initiated a program to investigate the torsional manifold of several isotopic species 
of methanol with one aim to create line lists with reliable frequencies and line strengths for 
astronomical observations and with the additional aim to analyze the intricate vibration-torsion-rotation 
interactions in their spectra. Our study on CD$_3$OH led to the detection of torsionally excited 
CD$_3$OH toward IRAS 16293$-$2422B \citep{CD3OH_rot_2022}. The following work on CD$_3$OD permitted 
the identification of some lines of this isotopolog in the same source, suggesting it may be observed 
unambiguously eventually \citep{CD3OD_rot_2023}. Our subsequent investigation of CH$_3$OD was, 
for example, instrumental for establishing the excitation conditions of the isotopolog toward 
IRAS 16293$-$2422B through emission lines involving higher energy levels because the low-energy 
transitions in $\varv_{\rm t} = 0$ were in part affected by opacity \citep{CH3OD_rot_2024}. 
The present work deals with a thorough study of the rotational spectrum of CH$_3$$^{17}$OH. 
We carried out extensive measurements in the lower millimeter region well into the submillimeter 
region to achieve a good coverage of experimental transition frequencies up to almost 1.1~THz. 
The new data were combined with previous microwave measurements \citep{CH3O-17-H_rot_1991} and 
in particular with published far-infrared measurements \citep{CH3O-17-H_FIR_MIR_2011}. 
We employed the resulting spectroscopic parameters to generate a line list in the ground 
and first excited torsional states of CH$_3$$^{17}$OH.

Given the high column densities of the hot molecular cores that are embedded in the high-mass 
star-forming region Sgr B2(N), it seemed natural to use our newly determined line list of 
CH$_3$$^{17}$OH to search for this isotopolog in the most sensitive spectral line survey that 
was carried out toward this region with ALMA so far, which is called Reexploring Molecular 
Complexity with ALMA \citep[ReMoCA,][]{ReMoCA_2019}.

The rest of the paper is organized as follows. Section~\ref{exptl} provides details 
on our laboratory measurements. The theoretical model, spectroscopic analysis, and 
fitting results are presented in Sects.~\ref{spec_backgr} and \ref{lab-results}. 
Section~\ref{astrosearch} describes our astronomical observations and the results 
of our search for CH$_3$$^{17}$OH, while Sect.~\ref{conclusion} gives the conclusions 
of our investigation.

\section{Experimental details}
\label{exptl}

All measurements were carried out at the Universit{\"a}t zu K{\"o}ln at room temperature 
employing a sample of methanol from Sigma-Aldrich Chemie GmbH that was enriched in $^{17}$O 
to 20\%; the remaining oxygen was mostly $^{16}$O, $^{18}$O was only marginally enriched. 
Three different spectrometers were used which all were equipped with 100~mm inner diameter
Pyrex glass cells of various lengths. The sample pressure was usually 1.5~Pa initially. 
The cells were pumped off and refilled after $\sim$8~h to $\sim$60~h because of pressure rise 
due to minute leaks. Lower pressures of $\sim$0.5~Pa were employed for remeasurements of 
selected individual lines showing $^{17}$O hyperfine splitting, while some weaker lines were 
recorded with a higher pressure of $\sim$5~Pa. Frequency multipliers usually from VDI and driven 
by Rohde \& Schwarz SMF~100A or Keysight E8257D synthesizers were utilized as sources 
above 70~GHz. An Agilent Technologies synthesizer E8257D was used for measurements of 
individual transitions between 38 and 69~GHz.

Two connected 7~m long absorption cells with Teflon windows were employed for the lowest frequency 
measurements in the 38$-$69~GHz region and those covering 70$-$130~GHz and 123$-$180~GHz. 
The frequency multiplier for this last frequency region was from RPG. 
The spectrometer system is similar to the one described by \citet{n-BuCN_rot_2012}. 
One 5~m long double-path cell equipped with Teflon windows was utilized for investigations in 
the regions 174$-$260~GHz, 260$-$362~GHz, and 370$-$510~GHz. Diode detectors were employed for 
the measurements described thus far. This second spectrometer system is identical to the one 
applied by \citet{OSSO_rot_2015}. A study of the rotational spectrum of 2-cyanobutane 
\citep{2-CAB_rot_2017} made also use of these two spectrometers and demonstrated that a frequency 
accuracy of 5~kHz can be achieved for a molecule with a much denser spectrum.

One 5~m long single path cell equipped with high-density polyethylene windows was employed for 
measurements at 360$-$504~GHz, 490$-$750~GHz, and 753$-$1095~GHz. The frequency multipliers were 
again from VDI, and a closed cycle liquid He-cooled InSb bolometer (QMC Instruments Ltd) was 
used as detector with this spectrometer system which is very similar to the one described by 
\citet{CH3SH_rot_2012}. We were able to achieve uncertainties of 10~kHz and even better with this 
spectrometer system for very symmetric lines with high signal-to-noise ratios (S/N) as demonstrated 
in recent studies on excited vibrational lines of CH$_3$CN \citep{MeCN_up2v4eq1_etc_2021} 
or on isotopic oxirane \citep{c-C2H4O_rot_2022,c-C2H3DO_rot_2023}. 
Uncertainties of 5~kHz, 10~kHz, 20~kHz, 30~kHz, 50~kHz, 100~kHz, and 200~kHz were assigned in 
the present study, depending on the symmetry of the line shape, the S/N, and on the frequency range. 
Frequency modulation was used with all spectrometer system, and the demodulation at $2f$ causes an 
isolated line to appear close to a second derivative of a Gaussian.

\section{Spectroscopic properties of CH$_3$$^{17}$OH and our theoretical approach}
\label{spec_backgr}

The methanol isotopolog CH$_3$$^{17}$OH is a nearly prolate top with $\kappa = (2B - A - C)/(A - C)$ 
being $-$0.9827. It displays a rather high coupling between internal and overall rotations in the 
molecule ($\rho \approx 0.809$) and a torsional potential barrier $V_3$ of about 374~cm$^{-1}$. 
The reduced barrier $s = 4V_3/9F$ is $\sim$6.04, where $F$ is the rotation constant of the internal 
rotor, putting its torsion problem to an intermediate barrier category \citep{RevModPhys.31.841}. 
Substitution of $^{16}$O by $^{18}$O in methanol changes the spectroscopic parameters only slightly, 
see, for example, \citet{CH3OH_rot_2008} and \citet{CH3O-18-H_rot_2007}, respectively. 
\citet{CH3O-17-H_FIR_MIR_2011} showed that the spectroscopic parameters of CH$_3$$^{17}$OH are 
quite close to the average of CH$_3$$^{16}$OH and CH$_3$$^{18}$OH, and a small and simple 
correction improves the estimates of the CH$_3$$^{17}$OH spectroscopic parameters.

\begin{figure}
\centering
   \includegraphics[width=9cm,angle=0]{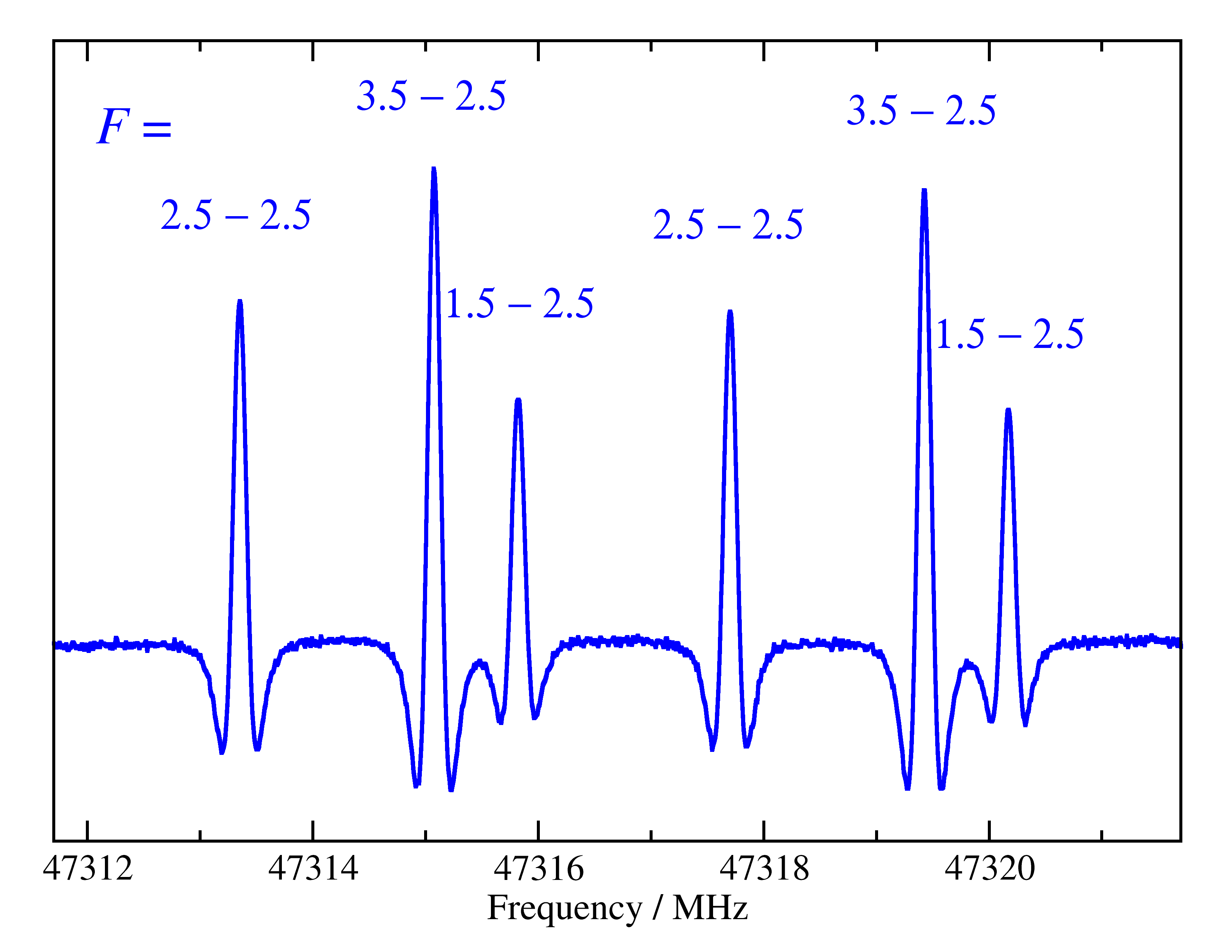}

\caption{Section of the rotational spectrum of CH$_3$$^{17}$OH displaying the 
$J_{K_a,K_c} = 1_{0,1} \leftarrow 0_{0,0}$ $a$-type transitions. The $\varv_{\rm t} = 0$ transition 
of A symmetry appears at lower frequencies, while the $\varv_{\rm t} = 0$ transition of E symmetry 
appears at higher frequencies. Both transitions are split into three HFS components whose 
upper and lower state $F$ quantum numbers are indicated, which are often designated as $F'$ 
and $F''$, respectively.}
\label{J=1-0-trans}
\end{figure}

The coupling between the hindered rotation of the methyl group and the overall rotational motion 
adds an oscillatory component to the energies of the $K$ levels, and this oscillatory component 
increases strongly upon torsional excitation, see for example Fig.~4 of \citet{CH3O-18-H_rot_2007}. 
This phenomenon is in part responsible for near-degeneracies of $K$ levels between torsional states 
already at low values of $\varv_{\rm t}$ that nourishes propagation of intervibrational interactions 
with non-torsional vibrations down to the lowest torsional states.

The dipole moment components of CH$_3$$^{17}$OH have, to the best of our knowledge, not been determined 
experimentally. However, accurate values of $\mu _a$ and $\mu _b$ were determined for CH$_3$$^{16}$OH 
in the ground vibrational as well as torsionally excited states up to $\varv_{\rm t} = 3$ by 
\citet{CH3OH_dip_2015}. The ground state values are $\mu _a = 0.8961$~(2)~D and $\mu _b = 1.4201$~(9)~D 
with an increase of almost 1\% upon excitation per torsional quantum. The ground state dipole moment 
components in the ground vibrational state of CH$_3$$^{18}$OH are marginally different from those 
of CH$_3$$^{16}$OH, namely $\mu _a = 0.8992$~(8)~D and $\mu _b = 1.4226$~(9)~D \citep{CH3O-18-H_dip_1996}. 
With the dipole moment components of CH$_3$$^{17}$OH most likely close to the average values of 
the two isotopologs, its rotational spectrum exhibits both fairly strong $a$- and $b$-type transitions. 
The dipole moment function of \citet{MEKHTIEV1999171} was employed in our calculations, where 
the values for the permanent dipole moment components of CH$_3$$^{16}$OH were replaced by averages 
of the CH$_3$$^{16}$OH and CH$_3$$^{18}$OH dipole moment components, namely $\mu_a = 0.89765$~D 
and $\mu_b = 1.42135$~D. The permanent dipole moment components were rotated from the principal axis 
system to the rho axis system of our Hamiltonian model.

\begin{figure}
\centering
   \includegraphics[width=9cm,angle=0]{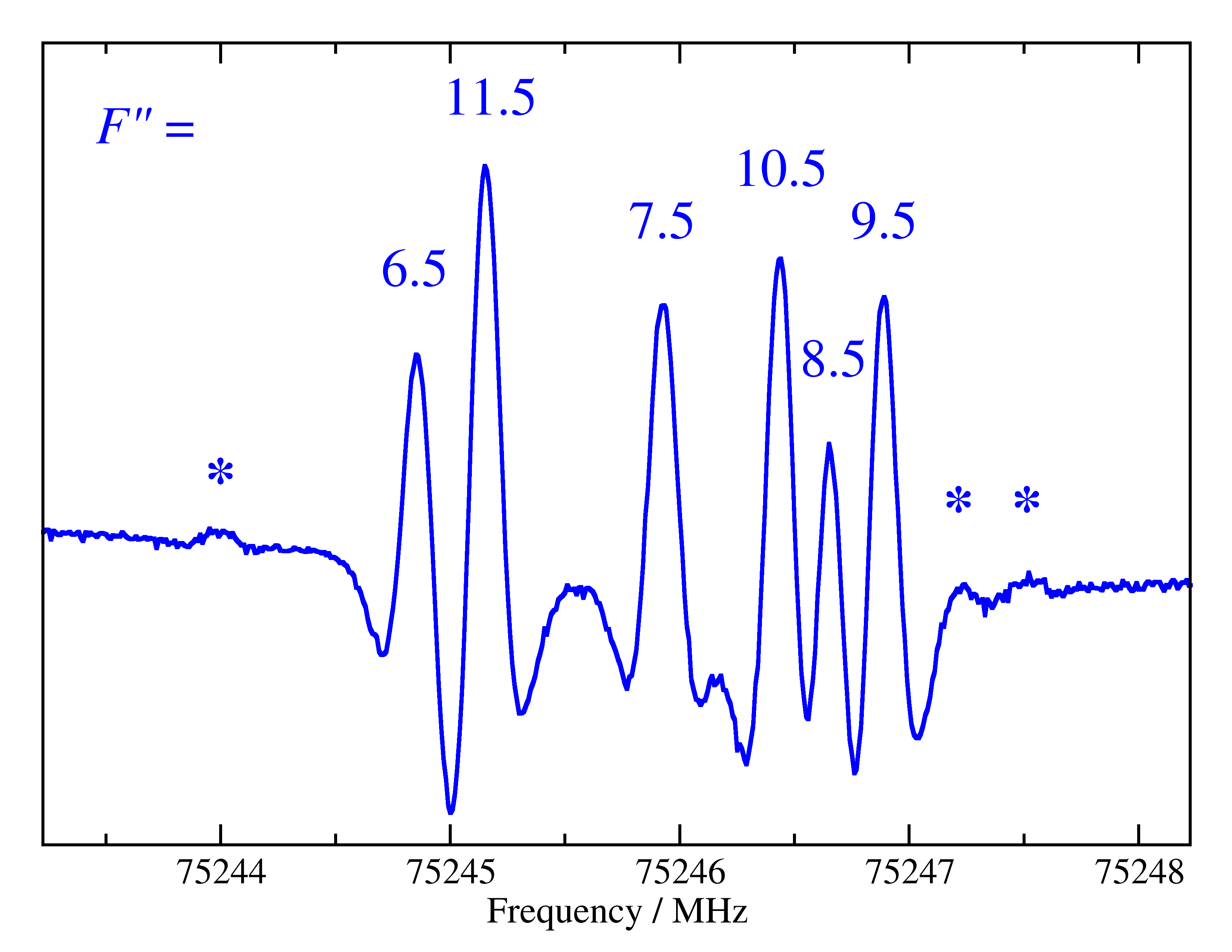}

\caption{Section of the rotational spectrum of CH$_3$$^{17}$OH showing the $\varv_{\rm t} = 0$ 
$J_{K_a,K_c} = 8_{2,6} \leftarrow 9_{1,8}$ transition of A symmetry with HFS splitting. The strong 
$\Delta F = \Delta J$ components are labeled with their lower state $F$ value; weaker and 
fairly well separated $\Delta F = 0$ components are labeled with asterisks.}
\label{HFS-trans}
\end{figure}

The nuclear spin $I = 5/2$ of $^{17}$O leads to hyperfine structure (HFS) splitting in the 
rotational spectrum of CH$_3$$^{17}$OH predominantly caused by its nuclear quadrupole moment. 
The rotational angular momentum {\bf J} is coupled with the spin angular momentum {\bf I} to yield 
the total angular momentum {\bf F}. The rotational selection rule $\Delta J = 0$, $\pm1$ is supplemented 
with the selection rule $\Delta F = 0$, $\pm1$. The $J = 1 \leftarrow 0$ transitions are split into 
three components, with the $\Delta F = 0$ component somewhat separated from the $\Delta F = \pm1$ 
components, as is shown in Fig.~\ref{J=1-0-trans} for the ground state $a$-type transitions of 
A and E symmetry, respectively. Energy levels with $J \ge I$ (here $J \ge 3$) are split 
into $2I + 1$ (six in our case) HFS levels with the strong HFS components being those with 
$\Delta F = \Delta J$ selection rules, which give rise to six strong components, as shown in 
Fig.~\ref{HFS-trans}. Components with $\Delta F \neq \Delta J$ are also allowed, but decrease 
rapidly with $J$ in intensity and thus are observable best for low values of $J$. Some of these 
weaker components are separated well enough from the stronger components in Fig.~\ref{HFS-trans}, 
such that they can be identified quite easily. 
It can also be seen in Fig.~\ref{HFS-trans}, that the $F = J \pm 2.5$ components occur fairly close 
in frequency and somewhat separated from the other two pairs with $F = J \pm 1.5$ and $F = J \pm 0.5$. 
This general appearance is quite typical for the HFS patterns of CH$_3$$^{17}$OH. 
The two components of each pair coalesce with decreasing splitting upon increase in $J$, thus, 
the HFS pattern collapse to three components. Upon further decrease of the splitting the 
components with $F = J \pm 1.5$ and $F = J \pm 0.5$ blend, which leads to a simplified pattern with 
roughly 1~:~2 intensity. Eventually, all six strong HFS components coalesce into one single line.

As in earlier investigations, we apply the rho-axis-method (RAM), which is named because of 
the choice of its axis system \citep{Hougen:1994}. In the rho-axis-method, the $z$ axis is 
coincident with the $\rho$ vector, which expresses the coupling between the angular momentum 
of the internal rotation $p_{\alpha}$ and that of the global rotation $J$. 
It is based on the work of \citet{Kirtman:1962}, \citet{CH_D3OH_D_rot_1968}, and 
\citet{Herbst:1984} and has proven to be a very effective approach in treating torsional 
large amplitude motions in methanol-like molecules. 
The RAM36 code \citep{Ilyushin:2010,Ilyushin:2013} was applied successfully for a number of 
near prolate tops with rather high $\rho$ and $J$ values, see for example \citet{Smirnov:2014}, 
\citet{CH3SH_rot_2019} and, in particular, for the CD$_3$OH isotopolog of methanol 
\citep{CD3OH_rot_2022} and for fluoral, CF$_3$CHO, \citep{Bermudez:2022}, which have $\rho$ 
values $\sim$0.89 and $\sim$0.92, respectively. 
In order to treat the HFS of CH$_3$$^{17}$OH, we resorted to the RAM36hf modification 
of the RAM36 code that was developed previously to analyze the rotational spectrum of 
N-methylformamide \citep{Belloche:2017} and was applied successfully to the analysis 
of methylarsine \citep{Motiyenko:2020}. 
The RAM36 and RAM36hf codes use the two-step diagonalization procedure of \citet{Herbst:1984}; 
in the present work, we kept 31 torsional basis functions at the first diagonalization step 
and 11 torsional basis functions at the second diagonalization step.

The energy levels are labeled in our fits and calculations by the free rotor quantum number $m$, 
the overall rotational angular momentum quantum number $J$, and a signed value of $K_a$, which 
is the axial $a$-component of the overall rotational angular momentum $J$. In the case of the A 
symmetry species, the $+/-$ sign of $K_a$ corresponds to the so-called "parity" designation, 
which is related to the $A_{1}/A_{2}$ symmetry species in the group $G_6$ \citep{Hougen:1994}. 
The signed value for the E symmetry species reflects the fact that the Coriolis-type interaction 
between the internal rotation and the global rotation causes $|K_a| > 0$ levels to split into 
a $K_a > 0$ level and a $K_a < 0$ level. We also provide $K_c$ values for convenience, 
but they are simply recalculated from the $J$ and $K_{a}$ values, $K_{c} = J - |K_{a}|$ for 
$K_{a} \geq 0$ and $K_{c} = J - |K_{a}| + 1$ for $K_{a} < 0$. The $m$ values 0, $-$3, and 3 
correspond to A symmetry levels of the $\varv_{\rm t} = 0$, 1, and 2 torsional states, 
respectively, while $m = 1$, $-$2, and 4 correspond to the respective E symmetry levels. 
The total angular momentum quantum number $F = J + I, J + I - 1, ..., |J - I|$ designates 
the HFS components. When the HFS components are not resolved, a hypothetical hyperfine 
free torsion-rotation transition is used in our fits, which is designated by $F= -1.0$.

\section{Spectroscopic results and discussion}
\label{lab-results}

Initial spectroscopic parameters of CH$_3$$^{17}$OH were evaluated by averaging the CH$_3$$^{16}$OH 
\citep{CH3OH_rot_2008} and CH$_3$$^{18}$OH \citep{CH3O-18-H_rot_2007} parameters of second and 
fourth orders. The quadrupole coupling parameters were taken from \citet{CH3O-17-H_rot_1991}. 
Subsequently, quadrupole coupling, rotational, and a few centrifugal distortion parameters 
plus the parameters $\rho$, $V_{3J}$, and $D_{3ac}$ were adjusted through fitting to the 
very sparse set of published transition frequencies \citep{CH3O-17-H_rot_1991}, which were 
the $J = 1 \leftarrow 0$ and $2 \leftarrow 1$ $a$-type transitions and the E symmetry 
$Q$-branch with $K = 2 \leftarrow 1$ with $2 \le J \le 23$. The resulting fit was 
the starting point of our present investigation.

\begin{figure*}
\centering
   \includegraphics[width=18cm,angle=0]{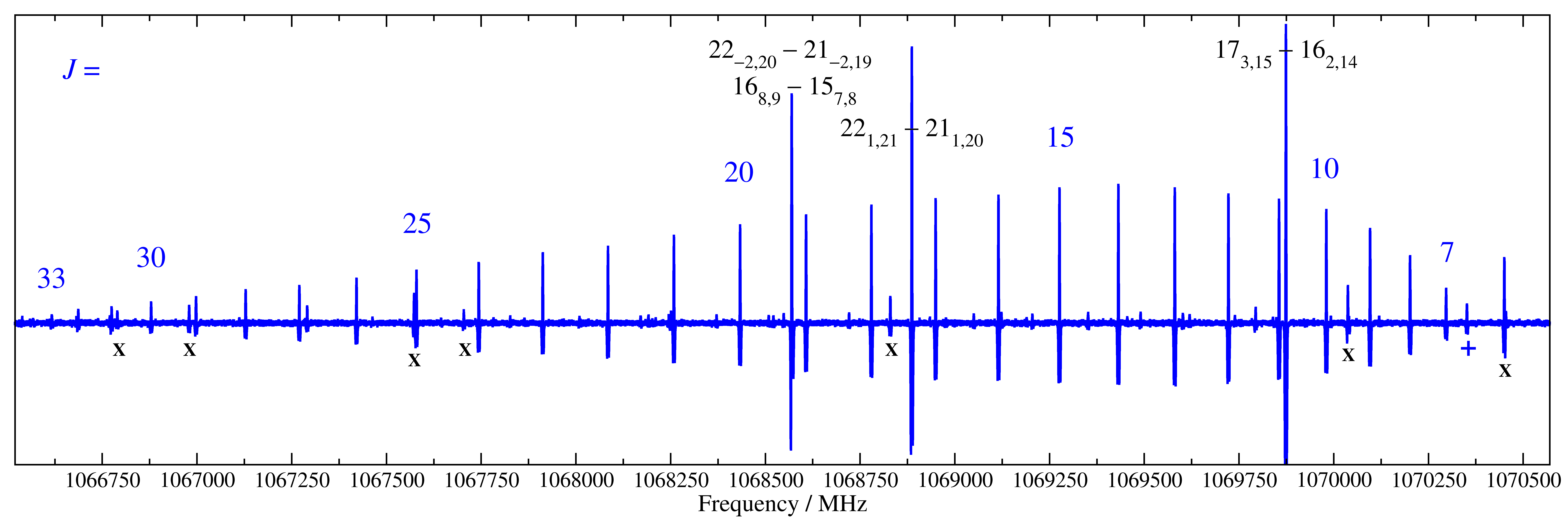}

\caption{Part of the terahertz spectrum of CH$_3$$^{17}$OH showing the A symmetry 
$\varv_{\rm t} = 0$ $K_a = 7 \leftarrow 6$ $Q$-branch. Selected $J$ values are indicated. 
Strong transitions of CH$_3$$^{16}$OH are indicated by there quantum numbers given 
in black. These are from left to right two blended $\varv_{\rm t} = 0$ and 1 transitions 
of E symmetry and two $\varv_{\rm t} = 0$ transitions of A symmetry. Weaker transitions of 
CH$_3$$^{16}$OH are indicated by black x, one weaker transition of CH$_3$$^{17}$OH is denoted 
by a blue plus sign. Several of the weak or very weak lines remain unassigned at present.}
\label{Q7-6}
\end{figure*}

The first round of assignments started from low-$J$ transitions and was straightforward for 
transitions whose HFS splitting depended mostly or only on $\chi _{aa}$, but difficult for 
those strongly affected by $\chi _{bb}$. This was hardly surprising, as \citet{CH3O-17-H_rot_1991} 
stated that their assignments do not constrain $\chi _{bb}$ well. 
We tried to evaluate the $^{17}$O quadrupole tensor of CH$_3$$^{17}$OH through a quantum chemical 
calculation with the program Gaussian~16 \citep{Gaussian16C} employing the B3LYP hybrid density 
functional \citep{Becke_1993,LYP_1988} with the aug-cc-pVTZ basis set \citep{cc-pVXZ_1989} of 
triple zeta quality. The calculated value $\chi _{aa} = -8.49$~MHz agreed very well with 
$-8.44 \pm 0.25$~MHz from the fit to the previous data \citep{CH3O-17-H_rot_1991}. Therefore, we 
kept $\chi _{bb}$ fixed to the calculated value of $-2.77$~MHz in the second fit, which included 
our first round assignments. These assignments included the $m = 1$ $0_{0,0} \leftarrow 1_{-1,1}$ 
transition along with the $m = 1$ $K_a = 1 \leftarrow 0$ $Q$-branch transitions with $1 \le J \le 8$. 
The calculated value of $\chi _{bb}$ turned out to be very helpful in the second round of assignments. 
Not only was it possible to know which transitions should display more or less well resolved HFS 
splitting, which transitions should have completely collapsed HFS patterns, and which ones are 
in between, it was also straightforward for a transition consisting of several lines to assign 
each absorption features to one or more HFS components in most cases. Moreover, the HFS patterns 
were often very useful in the assignment process.

\begin{figure}
\centering
   \includegraphics[width=9cm,angle=0]{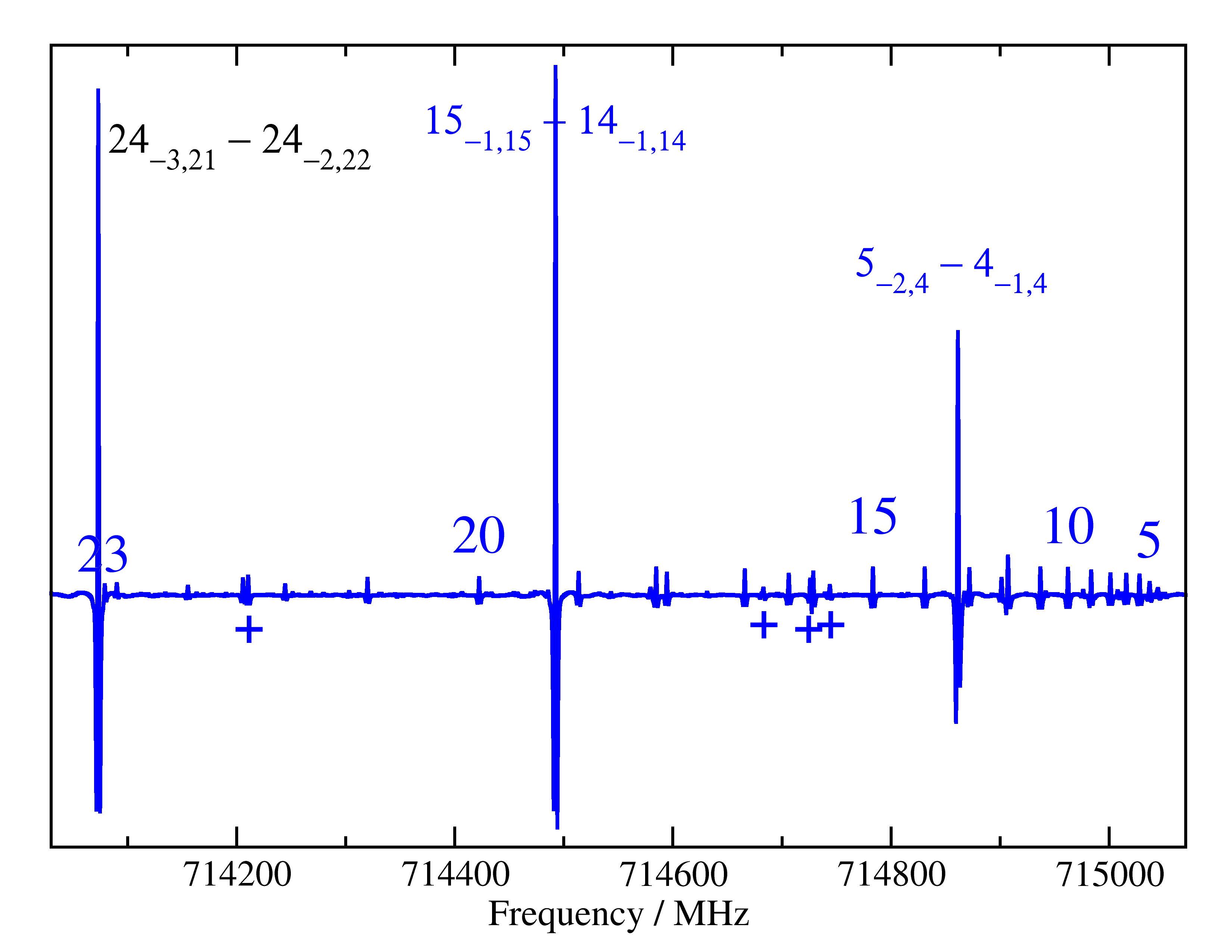}

\caption{Part of the terahertz spectrum of CH$_3$$^{17}$OH showing a large portion of the E 
symmetry $\varv_{\rm t} = 2$ $K_a = 3 \leftarrow 2$ $Q$-branch. Selected $J$ values are indicated. 
Two stronger A symmetry $\varv_{\rm t} = 0$ transitions are indicated by their quantum numbers 
given in blue. A strong E symmetry $\varv_{\rm t} = 0$ transition of CH$_3$$^{16}$OH is also 
indicated with quantum numbers in black. Four weaker transitions of CH$_3$$^{17}$OH are 
marked with blue plus signs.}
\label{vt2Q3-2}
\end{figure}

The assignments in the second round included several transitions with fairly large HFS splitting, 
which usually had low or somewhat low values of $J$. We also assigned extensively $a$-type $R$-branch 
transitions without HFS splitting up to $J = 14 \leftarrow 13$ and $K_a$ up to 8. Transitions involving 
$\Delta K_a = \pm1$ frequently showed larger deviations of more than 100~MHz even for small absolute 
values of $K_a$. Nevertheless, several transitions up to $J = 25$ and $K_a = 2$ could be assigned.

The second, quite extensive round of assignments laid the foundation for extensive additional 
line identifications. The third round already extended identifications of $a$-type $R$-branch 
transitions to the upper frequency limit of 1093~GHz, corresponding to $J = 23 \leftarrow 22$, 
and included some assignments of $\varv_{\rm t} = 1$. Assignments of $\Delta K_a = \pm1$ 
transitions were extended up to $J = 30$ and $K_a = 4$. Subsequent rounds of assignments enlarged 
the $J$ and $K_a$ range of $\varv_{\rm t} = 0$ and 1 transitions. Particularly useful in this context 
were $\Delta K_a  = \pm1$ $Q$-branches which frequently cover a large range of $J$ in a small frequency 
region, such as the $\varv_{\rm t} = 0$ $K_a = 7 \leftarrow 6$ branch shown in Fig.~\ref{Q7-6}. 
Care was taken to omit blended lines unless it was possible to account for the blending with 
transitions in the fit of CH$_3$$^{17}$OH.

We experienced difficulties in our previous investigations of CD$_3$OH \citep{CD3OH_rot_2022}, 
CD$_3$OD \citep{CD3OD_rot_2023}, and CH$_3$OD \citep{CH3OD_rot_2024} to fit data for the 
$\varv_{\rm t} = 2$ torsional state to within experimental uncertainties. We suspect these 
difficulties are caused by rovibrational interactions between small amplitude vibrations 
and higher ($\varv_{\rm t} \ge 3$) torsional states in the molecule. The effects of these 
interactions trickle down to $\varv_{\rm t} \le 2$ through torsion-torsion interactions. 
Since the emphasis of our first studies on various methanol isotopologs was and still is 
to provide line lists for radio astronomical observations, we limited our analyses largely 
to $\varv_{\rm t} = 0$ and 1 in all instances. The same approach was adopted here for 
CH$_3$$^{17}$OH. In order to account somewhat for $\varv_{\rm t} = 2$, we included 
transitions pertaining to the three lowest $K_a$ values of A and E symmetry in the fit, 
as in our previous studies of CD$_3$OH, CD$_3$OD, and CH$_3$OD, assuming that these 
$K$ levels are perturbed the least by the intervibrational interactions arising 
from low lying small amplitude vibrations. In the case of CH$_3$$^{17}$OH, 
these are $K = -1, 2, 3$ for the E species in $\varv_{\rm t} = 2$ and to 
$K = -4, 0, 4$ for the A species. Assignments of such $\varv_{\rm t} = 2$ $a$-type 
$R$-branch transitions were straightforward at a fairly late stage of the assignments. 
Subsequently, we also found several $\varv_{\rm t} = 2$ E symmetry transition between 
$K_a = 2$ and 3, notably a $Q$-branch with the origin near 715~GHz as shown in Fig.~\ref{vt2Q3-2}.

\begin{figure}
\centering
   \includegraphics[width=9cm,angle=0]{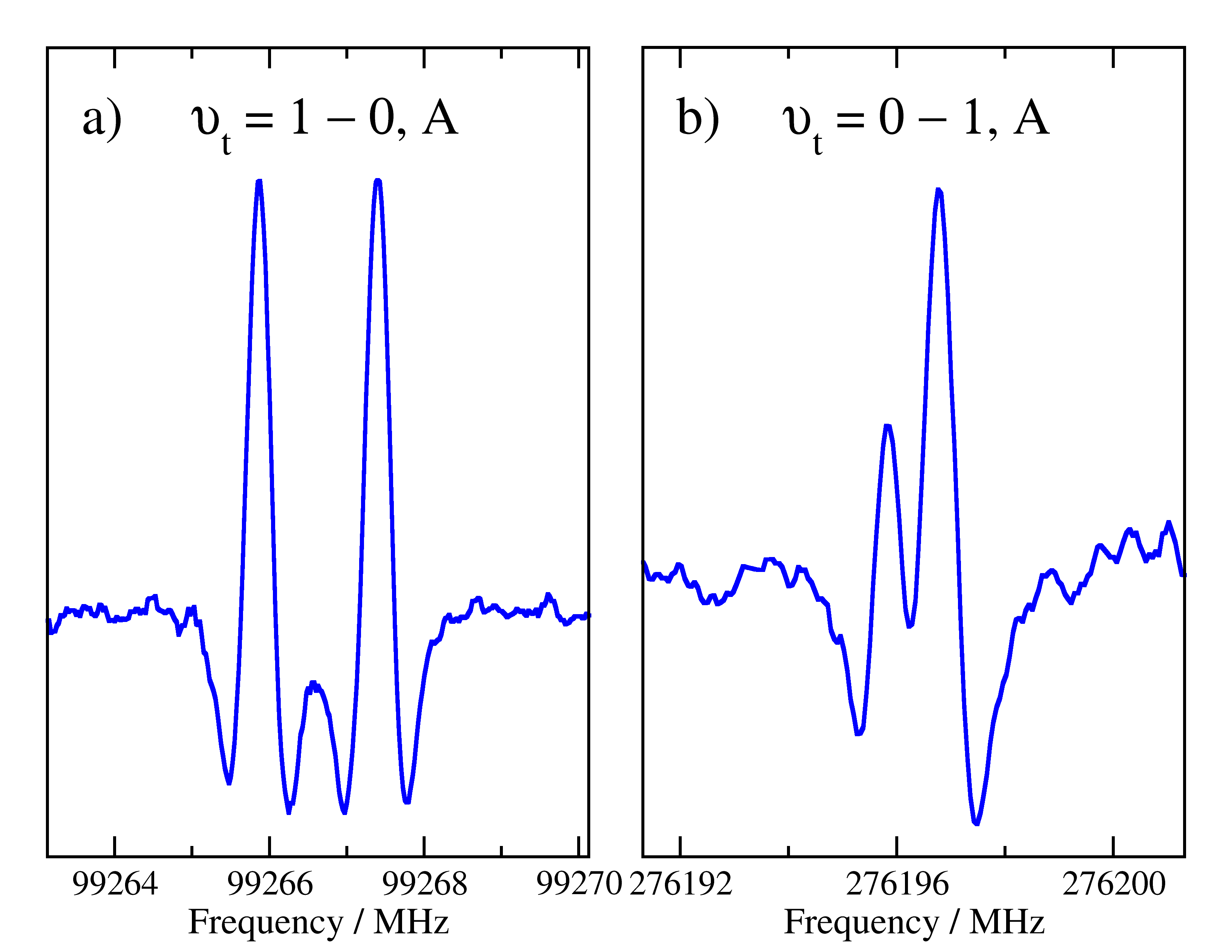}

\caption{Section of the millimeter-wave spectrum of CH$_3$$^{17}$OH displaying intertorsional  
transitions of A symmetry. Part a) shows the $\varv_{\rm t} = 1 \leftarrow 0$, 
$J_{K_a} = 24_{4} \leftarrow 25_{9}$ transitions with resolved asymmetry splitting. 
Part b) shows the $\varv_{\rm t} = 0 \leftarrow 1$, $J_{K_a} = 26_{9} \leftarrow 25_{6}$ 
transitions with unresolved asymmetry splitting, but partly resolved HFS splitting.}
\label{rovib-trans}
\end{figure}

In the same round, in which we made the first assignments of the $a$-type $R$-branch transitions 
in $\varv_{\rm t} = 2$, we also noted several A symmetry transitions between $\varv_{\rm t} = 0$ 
and 1, as shown exemplarily in Fig.~\ref{rovib-trans}. The transitions in this figure gain 
intensity for some $J$ because of the near-degeneracy of $\varv_{\rm t} = 0$ $K_a = 9$ of 
A symmetry with $\varv_{\rm t} = 1$ $K_a = 5$ also of A symmetry, which can be seen in Fig.~4 
of \citet{CH3O-18-H_rot_2007} for the case of CH$_3$$^{18}$OH. In the case of CH$_3$$^{17}$OH, 
these $K_a$ get particularly close for $J = 24$ and 25. This causes a mixing of 
the energy levels between these $K_a$ for the same $J$ around $J$ of 24 and 25. 
Transitions between $\varv_{\rm t} = 0$ $K_a = 9$ and $\varv_{\rm t} = 0$ $K_a = 8$ to 10 
transfer intensity to respective transitions between $\varv_{\rm t} = 1$ $K_a = 5$ and 
$\varv_{\rm t} = 0$ $K_a = 8$ to 10, and analogously, transitions between $\varv_{\rm t} = 1$ 
$K_a = 5$ and $\varv_{\rm t} = 1$ $K_a = 4$ to 6 transfer intensity to respective transitions 
between $\varv_{\rm t} = 0$ $K_a = 9$ and $\varv_{\rm t} = 1$ $K_a = 4$ to 6. 
All assigned rovibrational transitions of A symmetry determined with a microwave accuracy 
are caused by this $\varv_{\rm t} = 0$ $K_a = 9$ / $\varv_{\rm t} = 1$ $K_a = 5$ interaction. 
An interaction occurs also between $\varv_{\rm t} = 0$ $K_a = +9$ and $\varv_{\rm t} = 1$ 
$K_a = +3$ of E symmetry, albeit at higher $J$ of 33 and 34. Only one rovibrational transition 
associated with this perturbation was strong enough to be assigned with sufficient confidence.

Besides near-degeneracies between different $K_a$ of adjacent torsional states, there are also 
near-degeneracies within one torsional state, such as $K_a = -2$ and 4 at $J$ of 29 and 30 
as well as $K_a = -1$ and 3 at $J$ of 31 and 32 in the E species of $\varv_{\rm t} = 1$. 
But the effects of the first near-degeneracy are modest in terms of frequency and intensity 
alterations. The effects are more pronounced for the second near-degeneracy, as observed, 
for example, in the relatively weak $K_a = -1 \leftarrow 2$ $Q$-branch. The $J = 31$ transition 
is about a factor of four weaker than the $J = 28$ transition and shifted down by a few 100~MHz, 
whereas the $J = 32$ transition is about a factor of six stronger and shifted up by about 1~GHz 
with respect to the transitions with $J = 29$ and 33.

Despite the fact that our enriched methanol sample contains only 20\% $^{17}$O, which leads to a 
somewhat limited range of accessible $J$ and $K_a$ quantum numbers, we were able to assign microwave 
transitions up to $J = 45$ and $K_a = 16$. At the final stage, the FIR data from \citet{CH3O-17-H_FIR_MIR_2011} 
were added to the fit. Only transitions pertaining to the $\varv_{\rm t} = 0, 1$ torsional states and to 
the three lowest $K_a$ values of A and E symmetry in $\varv_{\rm t} = 2$ were added to the line list. 
Based on our fitting results and on an inspection of the 'smoothness' of spectroscopic branches, we have 
excluded from the fits 146 FIR lines out of 6677 line frequencies which satisfied our $\varv_{\rm t}$ 
cutoff criterion. In the majority of cases, the excluded FIR transitions correspond to the last transitions 
in $J$-series of $K_a' \leftarrow K_a''$ branches. Taking into account that our microwave dataset includes 
much higher $J$ values than the FIR dataset ($J_{\rm max}=32$), we assume that problems with those excluded 
"endings" in the $J$-series of different spectroscopic branches is an artifact of the Taylor-series expansions 
in powers of $J(J + 1)$ employed in \citet{CH3O-17-H_FIR_MIR_2011} for the spectral assignment.


\begin{table}
\begin{center}
\caption{Experimental nuclear quadrupole coupling parameters $\chi _{ij}$ (MHz) in comparison 
         to values from a B3LYP/aug-cc-pVTZ quantum-chemical calculation (QCC).}
\label{eQq-comparison}
\begin{tabular}[t]{lr@{}lr@{}l}
\hline \hline
Parameter  & \multicolumn{2}{c}{Exptl.} & \multicolumn{2}{c}{QCC} \\
\hline
$\chi _{aa}$              & $-$8&.2524~(54)    &  $-$8&.492       \\
$\chi _{bb}$              & $-$2&.852~(11)     &  $-$2&.774       \\
$\chi _{ab}$              & $-$3&.91~(44)      &  $-$4&.031       \\

\hline
\end{tabular}
\end{center}
\tablefoot{
Numbers in parentheses are one standard deviation in units of the least significant figures.
}
\end{table}
\begin{table*}[]
 \begin{center}
\caption{\label{tbl:statisticInf} Overview of the dataset and the fit quality }
\begin{tabular}{lrl|lrl}
\hline \hline
\multicolumn{3}{c|}{By measurement uncertainty}& \multicolumn{3}{c}{By torsional state} \\ 
\cline{1-6}
 
\multicolumn{1}{c}{Unc.$^a$} & \multicolumn{1}{c}{$\#^b$} & \multicolumn{1}{c|}{RMS$^c$} 
& \multicolumn{1}{c}{$\varv_{\rm t}^d$} & \multicolumn{1}{c}{$\#^b$} & \multicolumn{1}{c}{WRMS$^e$} \\

\cline{1-6}

0.005~MHz &  147 & 0.0049~MHz & $\varv_{\rm t}=0 \leftarrow 0$ & 3329 & 0.90 \\
0.010~MHz & 1003 & 0.0100~MHz & $\varv_{\rm t}=1 \leftarrow 1$ & 2731 & 0.90 \\
0.020~MHz &  890 & 0.0175~MHz & $\varv_{\rm t}=2 \leftarrow 2$ &  197 & 1.34 \\
0.030~MHz &  841 & 0.0279~MHz & $\varv_{\rm t}=1 \leftarrow 0$ & 5160 & 0.79 \\
0.050~MHz &  556 & 0.0483~MHz & $\varv_{\rm t}=2 \leftarrow 0$ &  507 & 0.89 \\
0.100~MHz &  304 & 0.0959~MHz & $\varv_{\rm t}=2 \leftarrow 1$ &  717 & 0.75 \\
0.200~MHz &    3 & 0.1366~MHz &  &  &  \\
$2\times 10^{-4}$~cm$^{-1}$ & 6473 & $1.6\times 10^{-4}$~cm$^{-1}$ &   &  & \\ 
$5\times 10^{-4}$~cm$^{-1}$ &   58 & $1.5\times 10^{-4}$~cm$^{-1}$ &   &  & \\ 
\hline

\end{tabular}
\end{center}
\tablefoot{$^{a}$ Estimated measurement uncertainties for each data group. $^{b}$ Number of lines (left part) 
or transitions (right part) of each category in the least-squares fit. Note that due to blending 10275 
measured line frequencies correspond to 12641  transitions in the fit.  $^{c}$ Root-mean-square (RMS) deviation 
of corresponding data group. $^{d}$ Upper and lower state torsional quantum number $\varv_{\rm t}$. 
$^{e}$ Weighted root-mean-square (WRMS) deviation of corresponding data group.}
\end{table*}

Our final CH$_3$$^{17}$OH dataset contains 6531 FIR and 3744 microwave line frequencies. 
Due to blending, these 10275 measured frequencies correspond to 12641 transitions with 
$J_{\rm max} = 45$ and $K_a \le 16$. A Hamiltonian model consisting of 121 parameters, 
of which three are quadrupole coupling parameters, provided a fit with a weighted root 
mean square (rms) deviation of 0.85, which was selected as our "best fit" for this paper. 
The 121 molecular parameters from our final fit are given in Table~\ref{tbl:ParametersTable}. 
Table~\ref{eQq-comparison} demonstrates that the experimental nuclear quadrupole coupling 
parameters and those from a quantum-chemical calculation agree well.
If we put aside quadrupole coupling parameters, the numbers of the terms in the model 
distributed between the orders $n_{\rm op} = 2$, 4, 6, 8, 10, 12 are 7, 22, 44, 34, 10, 1, 
respectively. This is consistent with the limits of determinable parameters of 7, 22, 50, 
95, 161, and 252 for these orders, as calculated from the differences between the total number 
of symmetry-allowed Hamiltonian terms of order $n_{\rm op}$ and the number of symmetry-allowed 
contact transformation terms of order $n_{\rm op} - 1$, when applying the ordering scheme 
of \citet{Nakagawa:1987}. The final set of the parameters converged perfectly in all 
three senses: (i) the relative change in the weighted rms deviation of the fit at the 
last iteration was about $\sim$$3 \times 10^{-7}$; (ii) the corrections to the parameter 
values generated at the last iteration are less than $\sim$10$^{-4}$ of the calculated 
parameter confidence intervals; (iii) the changes generated at the last iteration in 
the calculated frequencies are less than 1~kHz even for the FIR data. A summary of 
the quality of this fit is given in Table~\ref{tbl:statisticInf}. In the left part of 
Table~\ref{tbl:statisticInf}, the data are grouped by measurement uncertainty, and all 
data groups are fit within experimental uncertainties. We see the same good agreement in 
the right part of Table~\ref{tbl:statisticInf}, where the data are grouped by torsional state.

The final Hamiltonian model described above was used to calculate a CH$_3$$^{17}$OH line list 
in the ground and first excited torsional states for radio-astronomical observations. 
As in the cases of CD$_3$OH \citep{CD3OH_rot_2022}, CD$_3$OD \citep{CD3OD_rot_2023}, and 
CH$_3$OD \citep{CH3OD_rot_2024}, the list of CH$_3$$^{17}$OH transitions includes information 
on transition quantum numbers, transition frequencies, calculated uncertainties, lower state energies, 
and transition strengths. For convenience, we provide predictions for both hypothetical hyperfine 
free torsion-rotation transitions and torsion-rotation transitions with account of nuclear quadrupole 
hyperfine splittings. As already mentioned earlier, we label torsion-rotation levels by the free rotor 
quantum number $m$, the overall rotational angular momentum quantum number $J$, a signed value of 
$K_a$, and $K_c$. In the case of hyperfine structure the total angular momentum quantum number $F$ 
augments the labeling. To avoid unreliable extrapolations far beyond the quantum number coverage 
of the available experimental dataset, we limited our predictions to $\varv_{\rm t} \leq 1$, $J \leq 50$, 
and $|K_{a}| \leq 17$. The calculations were done from 1~GHz to 1.1~THz. Additionally, we limit our 
calculations to transitions for which calculated uncertainties are less than 0.1~MHz. The lower state 
energies are given referenced to the $J = 0$ A symmetry $\varv_{\rm t} = 0$ level. We provide 
additionally the torsion-rotation part of the partition function $Q_{\rm rt}$(T) of CH$_3$$^{17}$OH 
calculated from first principles, that is, via direct summation over the torsion-rotational levels. 
The maximum $J$ value is 65 for this calculation, and $n_{\varv_{\rm t}} = 11$ torsional states were 
taken into account. Due to the way the line strengths of hyperfine components are calculated 
(in the RAM36hf code we calculate line strength of a hypothetical hyperfine free torsion-rotation 
transition which then is multiplied by a relative intensity of a particular hyperfine component, 
with the sum of all relative intensities of all possible hyperfine components normalized to unity) 
the presented torsion-rotation part of the partition function should be used as it is both for 
calculation with and without hyperfine structure (in other words the $2I+1$ degeneracy due to 
quadrupole hyperfine splitting is already taken into account by the fact that the sum of 
relative intensities of hyperfine components is normalized to unity). The calculations, 
including the torsion-rotation part of the partition function, as well as the experimental 
line list from the present work, can be found in the online Supplementary material with this 
article\footnote{https://doi.org/10.5281/zenodo.12581728} and is also be available in the Cologne 
Database for Molecular Spectroscopy\footnote{https://cdms.astro.uni-koeln.de/classic/predictions/daten/Methanol/}, 
\citep[CDMS,][]{CDMS_2001,CDMS_2005,CDMS_2016}.

\section{Search for CH$_3$$^{17}$OH toward Sgr~B2(N)}
\label{astrosearch}

\begin{table*}[!ht]
 \begin{center}
 \caption{
 Parameters of our best-fit LTE model of methanol and its $^{13}$C, $^{18}$O, and $^{17}$O isotopologs toward Sgr~B2(N2b).
}
 \label{t:coldens}
 \vspace*{-1.2ex}
 \begin{tabular}{lcrcccccccr}
 \hline\hline
 \multicolumn{1}{c}{Molecule} & \multicolumn{1}{c}{Status\tablefootmark{(a)}} & \multicolumn{1}{c}{$N_{\rm det}$\tablefootmark{(b)}} & 
 \multicolumn{1}{c}{Size\tablefootmark{(c)}} & \multicolumn{1}{c}{$T_{\mathrm{rot}}$\tablefootmark{(d)}} & 
 \multicolumn{1}{c}{$N$\tablefootmark{(e)}} & \multicolumn{1}{c}{$F_{\rm vib}$\tablefootmark{(f)}} & 
 \multicolumn{1}{c}{$\Delta V$\tablefootmark{(g)}} & \multicolumn{1}{c}{$V_{\mathrm{off}}$\tablefootmark{(h)}} & 
 \multicolumn{1}{c}{$\frac{N_{\rm ref}}{N}$\tablefootmark{(i)}} \\ 
  & & & \multicolumn{1}{c}{\small ($''$)} & \multicolumn{1}{c}{\small (K)} & \multicolumn{1}{c}{\small (cm$^{-2}$)} & & 
 \multicolumn{1}{c}{\small (km~s$^{-1}$)} & \multicolumn{1}{c}{\small (km~s$^{-1}$)} & \\ 
 \hline
 CH$_3$OH\tablefootmark{(j)}$^\star$ & d & 69 &  0.5 &  140 &  8.0 (19) & 1.00 & 3.5 & $0.0$ &       1 \\ 
 $^{13}$CH$_3$OH, $\varv = 0$ & d & 30 &  0.5 &  140 &  3.2 (18) & 1.00 & 3.5 & $-0.2$ &      25 \\ 
 \hspace*{9.5ex} $\varv_{\rm t}=1$ & d & 10 &  0.5 &  140 &  3.2 (18) & 1.00 & 3.5 & $-0.2$ &      25 \\ 
 CH$_3$$^{18}$OH, $\varv = 0$ & d & 15 &  0.5 &  140 &  3.3 (17) & 1.00 & 3.5 & $-0.2$ &     240 \\ 
 \hspace*{9.5ex} $\varv_{\rm t}=1$ & t & 2 &  0.5 &  140 &  3.3 (17) & 1.00 & 3.5 & $-0.2$ &     240 \\ 
 CH$_3$$^{17}$OH, $\varv = 0$ & t & 2 &  0.5 &  140 &  1.0 (17) & 1.00 & 3.5 & $-0.2$ &     800 \\ 
\hline 
 \end{tabular}
 \end{center}
 \vspace*{-2.5ex}
 \tablefoot{
 \tablefoottext{a}{d: detection, t: tentative detection.}
 \tablefoottext{b}{Number of detected lines \citep[conservative estimate, see Sect.~3 of][]{EMoCA_with-D_2016}. 
  One line of a given species may mean a group of transitions of that species that are blended together.}
 \tablefoottext{c}{Source diameter (\textit{FWHM}).}
 \tablefoottext{d}{Rotational temperature.}
 \tablefoottext{e}{Total column density of the molecule. $x$ ($y$) means $x \times 10^y$. An identical value for all 
  listed torsional states of a molecule means that LTE is an adequate description of the torsional excitation.}
 \tablefoottext{f}{Correction factor that was applied to the column density to account for the contribution of 
  vibrationally excited states, in the cases where this contribution was not included in the partition function of 
  the spectroscopic predictions.}
 \tablefoottext{g}{Linewidth (\textit{FWHM}).}
 \tablefoottext{h}{Velocity offset with respect to the assumed systemic velocity of Sgr~B2(N2b), $V_{\mathrm{sys}} = 74.2$ km~s$^{-1}$.}
 \tablefoottext{i}{Column density ratio, with $N_{\rm ref}$ the column density of the previous reference species marked with a $\star$.}
 \tablefoottext{j}{The parameters were derived from the ReMoCA survey by \citet{Belloche22}.}
 }
 \end{table*}

We searched for interstellar CH$_3$$^{17}$OH toward the high-mass star-forming 
protocluster Sgr~B2(N) that is located in the Galactic center region at a 
distance of 8.2~kpc \citep[][]{Reid19}. We used the imaging spectral line 
survey Reexploring Molecular Complexity with ALMA (ReMoCA) that was performed 
toward Sgr~B2(N) with ALMA between 84.1 and 114.4~GHz with a spectral 
resolution of 488~kHz (1.7--1.3~km~s$^{-1}$), a median angular resolution of 
0.6$\arcsec$, and a median sensitivity of 0.8~mJy~beam$^{-1}$ (rms) 
corresponding to 0.27~K at 100~GHz. Details about the data reduction and the 
method of analysis of this survey can be found in \citet{ReMoCA_2019,Belloche22}. 
Here we analyzed the spectra toward the position Sgr~B2(N2b) that was defined 
by \citet{Belloche22} and is located in the secondary hot core Sgr~B2(N2) at 
($\alpha, \delta$)$_{\rm J2000}$= ($17^{\rm h}47^{\rm m}19{\fs}83, -28^\circ22'13{\farcs}6$). 
This position was chosen as a compromise between getting narrow line widths to reduce 
the level of spectral confusion and keeping a high enough H$_2$ column density 
to detect less abundant molecules.

\begin{table}
 \begin{center}
 \caption{
 Rotational temperature of the $^{13}$C, $^{18}$O, and $^{17}$O isotopologs of methanol derived 
 from their population diagrams toward Sgr~B2(N2b).
}
 \label{t:popfit}
 \vspace*{0.0ex}
 \begin{tabular}{lll}
 \hline\hline
 \multicolumn{1}{c}{Molecule} & \multicolumn{1}{c}{States\tablefootmark{(a)}} & 
 \multicolumn{1}{c}{$T_{\rm fit}$\tablefootmark{(b)}} \\ 
  & & \multicolumn{1}{c}{\small (K)} \\ 
 \hline
$^{13}$CH$_3$OH & $\varv=0$, $\varv_{\rm t}=1$ & 129.9 (3.2) \\ 
CH$_3$$^{18}$OH & $\varv=0$, $\varv_{\rm t}=1$ & 145.0 (7.5) \\ 
CH$_3$$^{17}$OH & $\varv=0$ & \hspace*{0.5ex}  161 (62) \\ 
\hline 
 \end{tabular}
 \end{center}
 \vspace*{-2.5ex}
 \tablefoot{
 \tablefoottext{a}{Torsional states that were taken into account to fit the population diagram.}
 \tablefoottext{b}{The standard deviation of the fit is given in parentheses. As explained in Sect.~3 
 of \citet{EMoCA_with-D_2016} and in Sect.~4.4 of \citet{ReMoCA_2019}, this uncertainty is purely 
statistical and should be viewed with caution. It may be underestimated.}
 }
 \end{table}

Before searching for CH$_3$$^{17}$OH toward Sgr~B2(N2b), we modeled the 
rotational emission of methanol and its $^{13}$C and $^{18}$O isotopologs under 
the assumption of local thermodynamic equilibrium (LTE) with the software Weeds 
\citep[][]{Maret11}. We used version 3 of the spectroscopic entry 32504 of 
methanol available in the CDMS \citep[][]{CDMS_2001,CDMS_2005,CDMS_2016}. 
The entry is based on \citet{CH3OH_rot_2008}; additional details are available in 
\citet{Belloche22}. To model the emission of the $^{13}$C and $^{18}$O isotopologs, 
we used version 2 of the CDMS entry 33502 and version 1 of the CDMS entry 34504, 
respectively. The entries were derived from \citet{12-13C-MeOH_1997} and 
\citet{CH3O-18-H_rot_2007}, respectively. Experimental transition frequencies 
in the range of our ALMA data are from \citet{13C-MeOH_rot_1986} and from 
the privately communicated methanol atlas of the Toyama University by 
\citet{MeOH_atlas_1995} in the case of $^{13}$CH$_3$OH and mainly from 
\citet{18O-MeOH_rot_1996} and \citet{18O-MeOH_rot_1998} in the case of 
CH$_3$$^{18}$OH.

The results of our LTE modeling of the emission of the main isotopolog of 
methanol toward Sgr~B2(N2b) were published in \citet{Belloche22} and the 
derived LTE parameters are recalled in Table~\ref{t:coldens}. We assumed the 
same parameters for the LTE modeling of the $^{13}$C and $^{18}$O isotopologs, 
keeping only their column densities as free parameters. Both isotopologs are 
clearly detected in their torsional ground state and their first torsionally 
excited state (see 
Figs.~\ref{f:remoca_ch3oh_13c_ve0_n2b}--\ref{f:remoca_ch3oh_18o_ve1_n2b}).
Figures~\ref{f:popdiag_ch3oh_13c_n2b} and \ref{f:popdiag_ch3oh_18o_n2b} show 
their population diagrams that were constructed using transitions that are not
too heavily contaminated by the contribution of other molecules. A fit to each 
population diagram yields a rotational temperature of $130 \pm 3$~K and 
$145 \pm 7$~K, respectively (see Table~\ref{t:popfit}). Both values are 
consistent with the temperature derived for methanol toward Sgr~B2(N2b) 
\citep[$138 \pm 2$~K, see][]{Belloche22} and with the temperature of 
140~K assumed for the LTE modeling of all methanol isotopologs. The best-fit 
column densities derived for $^{13}$CH$_3$OH and CH$_3$$^{18}$OH are reported 
in Table~\ref{t:coldens}. The red spectra in 
Figs.~\ref{f:remoca_ch3oh_13c_ve0_n2b}--\ref{f:remoca_ch3oh_18o_ve1_n2b} were
computed with these column densities. The column densities listed in 
Table~\ref{t:coldens} yield isotopic ratios $^{12}$C/$^{13}$C = 25 and 
$^{16}$O/$^{18}$O = 240, which are consistent with the values expected 
for the Galactic center region \citep[][]{isotopic_ratios_1994}.

\begin{figure*}
\centerline{\resizebox{0.85\hsize}{!}{\includegraphics[angle=0]{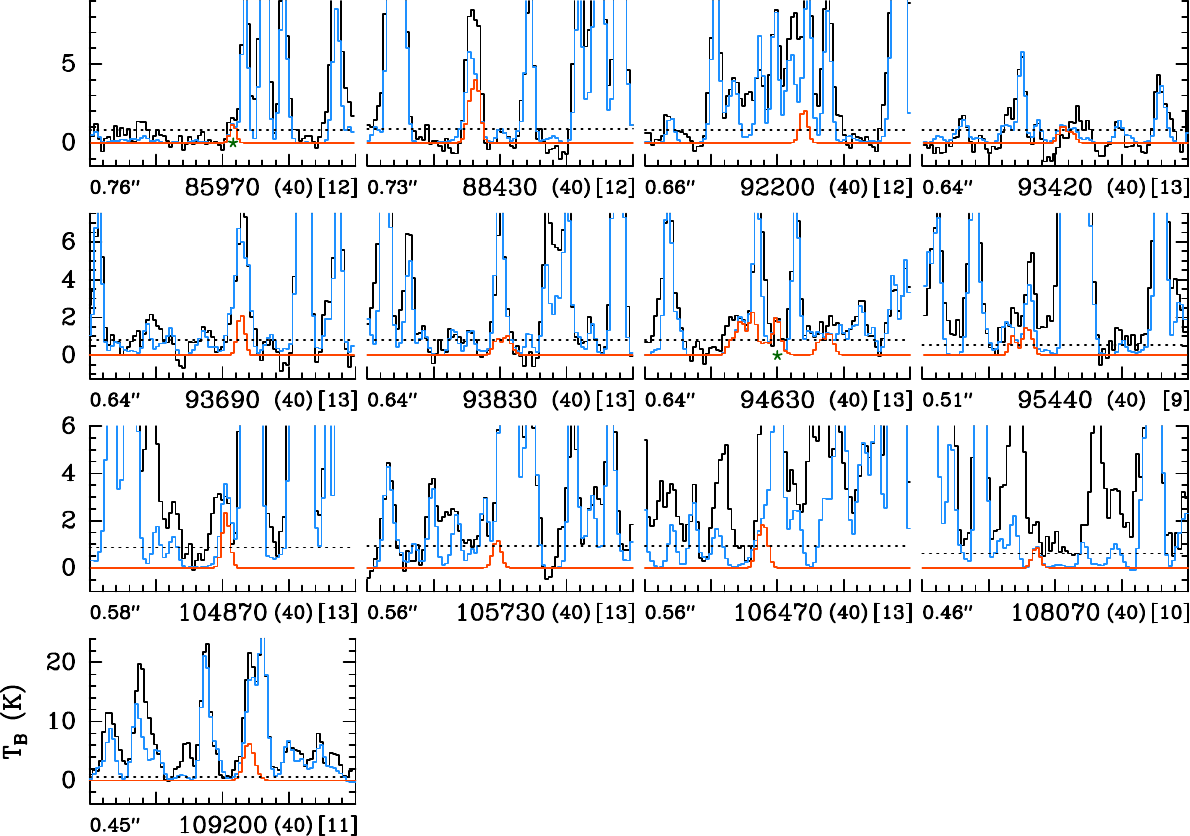}}}
\caption{Selection of rotational transitions of CH$_3$$^{17}$OH covered by the 
ReMoCA survey. The LTE synthetic spectrum of CH$_3$$^{17}$OH used to derive 
its column density toward Sgr~B2(N2b) is displayed in red and overlaid on the 
observed spectrum shown in black. The blue synthetic spectrum contains the 
contributions of all molecules identified in our survey so far, including 
CH$_3$$^{17}$OH. The values written below each panel correspond from left 
to right to the half-power beam width, the central frequency in MHz, the width 
in MHz of each panel in parentheses, and the continuum level in K of the 
baseline-subtracted spectra in brackets. The y-axis is labeled in brightness 
temperature units (K). The dotted line indicates the $3\sigma$ noise level.}
\label{f:remoca_ch3oh_17o_n2b}
\end{figure*}

Using the LTE parameters obtained for methanol and its $^{13}$C and $^{18}$O
isotopologs, we computed synthetic spectra for CH$_3$$^{17}$OH and searched for
matches in the ReMoCA spectrum of Sgr~B2(N2b). Our best-fit model is shown in 
red in Fig.~\ref{f:remoca_ch3oh_17o_n2b} and the corresponding column density 
is given in Table~\ref{t:coldens}. Two rotational lines of CH$_3$$^{17}$OH are
sufficiently free of contamination and bright enough to be categorized as
detected. They are marked with green stars in 
Fig.~\ref{f:remoca_ch3oh_17o_n2b}. The other lines are to some extent 
contaminated by emission from other molecules, but some of them can still
be used to construct a population diagram because we can subtract the 
contamination from their measured integrated intensities by using our LTE 
model that contains the contribution of all molecules identified so far. A fit
to the population diagram shown in Fig.~\ref{f:popdiag_ch3oh_17o_n2b} yields
a temperature of $160 \pm 60$~K (Table~\ref{t:popfit}), which is not well
constrained but is consistent with the temperatures measured for the other
methanol isotopologs.

The column density obtained for CH$_3$$^{17}$OH implies isotopic ratios 
$^{16}$O/$^{17}$O = 800 and $^{18}$O/$^{17}$O = 3.3, with uncertainties on the 
order of 20\% due to the moderate signal-to-noise ratios of the CH$_3$$^{17}$OH
lines. The $^{18}$O/$^{17}$O isotopic ratio is consistent with values reported 
earlier for other molecules in Sgr~B2. \citet{Wouterloot08} obtained a 
$^{18}$O/$^{17}$O ratio of $2.9 \pm 0.1$ for Sgr~B2(M) from a large velocity 
gradient (LVG) analysis of single-dish observations of CO isotopologs in the 
1--0, 2--1,and 3--2 transitions. \citet{18O-17O_Gal_ratio_2020} derived an 
average ratio of $3.15 \pm 0.08$ over a $10\arcmin \times 10\arcmin$ single-dish 
map of the Sgr~B2 molecular cloud in the 1--0 transition of C$^{18}$O and C$^{17}$O,
and a value of $3.25 \pm 0.42$ from a deeper integration toward a position 
a bit less than $1\arcmin$ north of Sgr~B2(N). \citet{Neill14} derived 
an isotopic ratio of $3.2 \pm 0.4$ from far-infrared observations of 
H$_2$$^{18}$O and H$_2$$^{17}$O detected in absorption in the envelope 
of Sgr B2(N) with the Herschel satellite. All these ratios are consistent
with the value of $3.1 \pm 0.6$ derived four decades ago by \citet{Guelin82} 
from single-dish measurements performed toward Sgr~B2 in the 1--0 transition 
of HC$^{18}$O$^+$ and HC$^{17}$O$^+$ with an angular resolution of about 
2$\arcmin$. The excellent agreement between the $^{18}$O/$^{17}$O isotopic 
ratio that we derived for methanol and the ratios reported in the past for 
other molecules in Sgr B2 gives strong support to our tentative interstellar 
identification of CH$_3$$^{17}$OH toward Sgr~B2(N2b).

\section{Conclusion}
\label{conclusion}

An extensive study of the torsion-rotation spectrum of CH$_3$$^{17}$OH was carried out in the broad 
frequency range from 38~GHz to 1.1~THz. Transitions involving the $\varv_{\rm t}$ = 0, 1, and 2 
torsional states with $J$ up to 45 and $K_a$ up to 16 were assigned and analyzed in the current work 
employing a torsion-rotation RAM Hamiltonian. We achieved a fit within the experimental uncertainties 
with a weighted rms deviation of 0.85 for the dataset, which consists of 6531 FIR and 3744 microwave 
line frequencies. On the basis of the obtained Hamiltonian model, a line list in the ground and 
first excited torsional states of the CH$_3$$^{17}$OH molecule was calculated to facilitate 
radio-astronomical searches of this methanol isotopolog.

We report a tentative detection of CH$_3$$^{17}$OH with ALMA toward Sgr~B2(N2b) along 
with detections of the more abundant isotopologs of methanol. The measurements yield 
isotopic ratios $^{12}$C/$^{13}$C = 25, $^{16}$O/$^{18}$O = 240, and $^{18}$O/$^{17}$O = 3.3. 
The latter value is consistent with isotopic ratios reported earlier for CO, HCO$^+$, and 
H$_2$O toward Sgr~B2, which strongly supports our tentative interstellar identification 
of CH$_3$$^{17}$OH.


\begin{acknowledgements}
We acknowledge support by the Deutsche Forschungsgemeinschaft via the collaborative research 
centers SFB~956 (project ID 184018867) subproject B3 and SFB~1601 (project ID 500700252) 
subprojects A4 and Inf as well as the Ger{\"a}tezentrum SCHL~341/15-1 
(``Cologne Center for Terahertz Spectroscopy''). 
Part of the research in Kharkiv was carried out under support of the Volkswagen foundation. 
The assistance of the Science and Technology Center in the Ukraine is acknowledged 
(STCU partner project P756). V.V.I. acknowledges financial support from Deutsche 
Forschungsgemeinschaft (grant number BA2176/9-1). 
Our research benefited from NASA's Astrophysics Data System (ADS). 
This paper makes use of the following ALMA data: ADS/JAO.ALMA \# 2016.1.00074.S. 
ALMA is a partnership of ESO (representing its member states), NSF (USA) and NINS (Japan), 
together with NRC (Canada), MOST and ASIAA (Taiwan), and KASI (Republic of Korea), 
in cooperation with the Republic of Chile. The Joint ALMA Observatory is operated by ESO, 
AUI/NRAO and NAOJ.
\end{acknowledgements}


\bibliographystyle{aa} 
\bibliography{CH3O-17-H}

\onecolumn
\begin{appendix}
\label{appendix}

\section{Spectroscopic parameters of the RAM Hamiltonian for the CH$_3$$^{17}$OH molecule}
\label{spec-parameters}


\begin{longtable}{lllr}
\caption{\label{tbl:ParametersTable} Fitted parameters of the RAM Hamiltonian for the CH$_3$$^{17}$OH molecule}\\

\hline\hline 
$n_{tr}$\textit{$^a$} & Par.\textit{$^{b}$} & Operator\textit{$^c$} & Value\textit{$^{d,e}$} \\
\hline
\endfirsthead
\caption{continued.}\\
\hline\hline $n_{tr}$\textit{$^a$} & Par.\textit{$^{b}$} & Operator\textit{$^c$} & Value\textit{$^{d,e}$} \\
\hline
\endhead
\hline

$  2_{ 2, 0}$ &    $(1/2)V_3$  &   $(1-\cos 3\alpha)$                                         &  $   186.94207(64) $ \\ 
$  2_{ 2, 0}$ &    $F       $  &   $p_\alpha^2$                                               &  $  27.5312352(20) $ \\ 
$  2_{ 1, 1}$ &    $\rho     $  &   $P_ap_\alpha$                                              &  $ 0.8094036988(18) $ \\ 
$  2_{ 0, 2}$ &    $A_{\rm RAM}       $  &   $P_a^2$                                                    &  $     4.22572(28) $ \\ 
$  2_{ 0, 2}$ &    $B_{\rm RAM}       $  &   $P_b^2$                                                    &  $    0.8051588(47) $ \\ 
$  2_{ 0, 2}$ &    $C_{\rm RAM}       $  &   $P_c^2$                                                    &  $    0.7753620(35) $ \\ 
$  2_{ 0, 2}$ &    $2D_{ab} $  &   $(1/2)\{P_a{,}P_b\}$                                       &  $    -0.0108380(64) $ \\ 
$  4_{ 4, 0}$ &    $(1/2)V_6$  &   $(1-\cos 6\alpha)$                                         &  $      -0.8414(44) $ \\ 
$  4_{ 4, 0}$ &    $F_m      $  &   $p_\alpha^4$                                               &  $   -0.8818018(89) \times 10^{ -2}$ \\ 
$  4_{ 3, 1}$ &    $\rho_m    $  &   $P_ap_\alpha^3$                                            &  $   -0.3441336(28) \times 10^{ -1}$ \\ 
$  4_{ 2, 2}$ &    $V_{3J}     $  &   $P^2(1-\cos 3\alpha)$                                      &  $    -0.233190(20) \times 10^{ -2}$ \\ 
$  4_{ 2, 2}$ &    $V_{3K}     $  &   $P_a^2(1-\cos 3\alpha)$                                    &  $      0.12153(13) \times 10^{ -1}$ \\ 
$  4_{ 2, 2}$ &    $V_{3bc}    $  &   $(P_b^2-P_c^2)(1-\cos 3\alpha)$                            &  $      -0.4030(86) \times 10^{ -4}$ \\ 
$  4_{ 2, 2}$ &    $V_{3ab}   $  &   $(1/2)\{P_a{,}P_b\}(1-\cos 3\alpha)$                       &  $     0.178794(29) \times 10^{ -1}$ \\ 
$  4_{ 2, 2}$ &    $F_J      $  &   $P^2p_\alpha^2$                                            &  $  -0.11533011(92) \times 10^{ -3}$ \\ 
$  4_{ 2, 2}$ &    $F_K      $  &   $P_a^2p_\alpha^2$                                          &  $   -0.5093083(33) \times 10^{ -1}$ \\ 
$  4_{ 2, 2}$ &    $F_{bc}     $  &   $(P_b^2-P_c^2)p_\alpha^2$                                  &  $      -0.1251(12) \times 10^{ -3}$ \\ 
$  4_{ 2, 2}$ &    $F_{ab}     $  &   $(1/2)\{P_a{,}P_b\}p_\alpha^2$                             &  $       0.1538(16) \times 10^{ -5}$ \\ 
$  4_{ 2, 2}$ &    $D_{3ac}    $  &   $(1/2)\{P_a{,}P_c\}\sin 3\alpha$                           &  $     0.256158(44) \times 10^{ -1}$ \\ 
$  4_{ 2, 2}$ &    $D_{3bc}   $  &   $(1/2)\{P_b{,}P_c\}\sin 3\alpha$                           &  $       -0.138(16) \times 10^{ -3}$ \\ 
$  4_{ 1, 3}$ &    $\rho_J    $  &   $P^2P_ap_\alpha$                                           &  $   -0.1856334(19) \times 10^{ -3}$ \\ 
$  4_{ 1, 3}$ &    $\rho_K   $  &   $P_a^3p_\alpha$                                            &  $   -0.3354061(19) \times 10^{ -1}$ \\ 
$  4_{ 1, 3}$ &    $\rho_{bc}   $  &   $(1/2)\{P_a{,}(P_b^2-P_c^2)\}p_\alpha$                     &  $      -0.2183(12) \times 10^{ -3}$ \\ 
$  4_{ 0, 4}$ &    $-\Delta_J     $  &   $P^4$                                                      &  $   -0.1626985(20) \times 10^{ -5}$ \\ 
$  4_{ 0, 4}$ &    $-\Delta_{JK} $  &   $P^2P_a^2$                                                 &  $     -0.10148(27) \times 10^{ -3}$ \\ 
$  4_{ 0, 4}$ &    $-\Delta_K     $  &   $P_a^4$                                                    &  $   -0.8286520(49) \times 10^{ -2}$ \\ 
$  4_{ 0, 4}$ &    $-2\delta_J$  &   $P^2(P_b^2-P_c^2)$                                         &  $   -0.1107985(58) \times 10^{ -6}$ \\ 
$  4_{ 0, 4}$ &    $-2\delta_K$  &   $(1/2)\{P_a^2{,}(P_b^2-P_c^2)\}$                           &  $     -0.91153(87) \times 10^{ -4}$ \\ 
$  4_{ 0, 4}$ &    $D_{abJ}    $  &   $(1/2)P^2\{P_a{,}P_b\}$                                    &  $     -0.16976(77) \times 10^{ -6}$ \\ 
$  6_{ 6, 0}$ &    $(1/2)V_9 $  &   $(1-\cos 9\alpha)$                                         &  $        0.786(17)$ \\ 
$  6_{ 6, 0}$ &    $F_{mm} $  &   $p_\alpha^6$                                               &  $      0.10069(20) \times 10^{ -4}$ \\ 
$  6_{ 5, 1}$ &    $\rho_{mm} $  &   $P_ap_\alpha^5$                                            &  $      0.67059(99) \times 10^{ -4}$ \\ 
$  6_{ 4, 2}$ &    $V_{6J} $  &   $P^2(1-\cos 6\alpha)$                                      &  $       -0.456(14) \times 10^{ -4}$ \\ 
$  6_{ 4, 2}$ &    $V_{6K} $  &   $P_a^2(1-\cos 6\alpha)$                                    &  $      -0.8042(92) \times 10^{ -2}$ \\ 
$  6_{ 4, 2}$ &    $V_{6bc} $  &   $(P_b^2-P_c^2)(1-\cos 6\alpha)$                            &  $      -0.2379(12) \times 10^{ -4}$ \\ 
$  6_{ 4, 2}$ &    $V_{6ab} $  &   $(1/2)\{P_a{,}P_b\}(1-\cos 6\alpha)$                       &  $      -0.1611(91) \times 10^{ -3}$ \\ 
$  6_{ 4, 2}$ &    $F_{mJ} $  &   $P^2p_\alpha^4$                                            &  $       0.8956(10) \times 10^{ -7}$ \\ 
$  6_{ 4, 2}$ &    $F_{mK}  $  &   $P_a^2p_\alpha^4$                                          &  $      0.18102(20) \times 10^{ -3}$ \\ 
$  6_{ 4, 2}$ &    $F_{mab}  $  &   $(1/2)\{P_a{,}P_b\}p_\alpha^4$                             &  $       -0.250(11) \times 10^{ -8}$ \\ 
$  6_{ 4, 2}$ &    $D_{6ac}   $  &   $(1/2)\{P_a{,}P_c\}\sin 6\alpha$                           &  $      -0.3458(40) \times 10^{ -2}$ \\ 
$  6_{ 3, 3}$ &    $\rho_{mJ}  $  &   $P^2P_ap_\alpha^3$                                         &  $      0.34612(28) \times 10^{ -6}$ \\ 
$  6_{ 3, 3}$ &    $\rho_{mK} $  &   $P_a^3p_\alpha^3$                                          &  $      0.25554(22) \times 10^{ -3}$ \\ 
$  6_{ 3, 3}$ &    $\rho_{3bc}$  &   $(1/2)\{P_a{,}P_b{,}P_c{,}p_\alpha{,}\sin 3\alpha\}$       &  $        0.412(11) \times 10^{ -4}$ \\ 
$  6_{ 2, 4}$ &    $V_{3JJ} $  &   $P^4(1-\cos 3\alpha)$                                      &  $      0.10480(28) \times 10^{ -7}$ \\ 
$  6_{ 2, 4}$ &    $V_{3JK} $  &   $P^2P_a^2(1-\cos 3\alpha)$                                 &  $      -0.3861(21) \times 10^{ -6}$ \\ 
$  6_{ 2, 4}$ &    $V_{3KK} $  &   $P_a^4(1-\cos 3\alpha)$                                    &  $       0.6713(36) \times 10^{ -6}$ \\ 
$  6_{ 2, 4}$ &    $V_{3bcJ} $  &   $P^2(P_b^2-P_c^2)(1-\cos 3\alpha)$                         &  $      0.24337(62) \times 10^{ -8}$ \\ 
$  6_{ 2, 4}$ &    $V_{3bcK} $  &   $(1/2)\{P_a^2{,}(P_b^2-P_c^2)\}(1-\cos 3\alpha)$           &  $      -0.2185(64) \times 10^{ -5}$ \\ 
$  6_{ 2, 4}$ &    $V_{3b2c2} $  &   $(1/2)\{P_b^2{,}P_c^2\}\cos 3\alpha$                       &  $       0.4618(62) \times 10^{ -7}$ \\ 
$  6_{ 2, 4}$ &    $V_{3abJ} $  &   $(1/2)P^2\{P_a{,}P_b\}(1-\cos 3\alpha)$                    &  $      -0.2421(10) \times 10^{ -6}$ \\ 
$  6_{ 2, 4}$ &    $V_{3abK}$  &   $(1/2)\{P_a^3{,}P_b\}(1-\cos 3\alpha)$                     &  $      -0.1439(16) \times 10^{ -5}$ \\ 
$  6_{ 2, 4}$ &    $F_{JJ}     $  &   $P^4p_\alpha^2$                                            &  $       0.5256(26) \times 10^{ -9}$ \\ 
$  6_{ 2, 4}$ &    $F_{JK}   $  &   $P^2P_a^2p_\alpha^2$                                       &  $      0.51158(30) \times 10^{ -6}$ \\ 
$  6_{ 2, 4}$ &    $F_{KK} $  &   $P_a^4p_\alpha^2$                                          &  $      0.20008(14) \times 10^{ -3}$ \\ 
$  6_{ 2, 4}$ &    $D_{3acJ}  $  &   $(1/2)P^2\{P_a{,}P_c\}\sin 3\alpha$                        &  $      -0.5868(21) \times 10^{ -6}$ \\ 
$  6_{ 2, 4}$ &    $D_{3acK} $  &   $(1/2)\{P_a^3{,}P_c\}\sin 3\alpha$                         &  $      -0.1263(22) \times 10^{ -5}$ \\ 
$  6_{ 2, 4}$ &    $D_{3bcJ}  $  &   $(1/2)P^2\{P_b{,}P_c\}\sin 3\alpha$                        &  $      -0.1123(17) \times 10^{ -7}$ \\ 
$  6_{ 2, 4}$ &    $D_{3bcK} $  &   $(1/2)\{P_a^2{,}P_b{,}P_c\}\sin 3\alpha$                   &  $       0.2956(79) \times 10^{ -4}$ \\ 
$  6_{ 2, 4}$ &    $D_{3acbc}$  &   $(1/2)(\{P_a{,}P_b^2{,}P_c\}-\{P_a{,}P_c^3\})\sin 3\alpha$ &  $      -0.3816(13) \times 10^{ -6}$ \\ 
$  6_{ 2, 4}$ &    $D_{3bcbc} $  &   $(1/2)(\{P_b^3{,}P_c\}-\{P_b{,}P_c^3\})\sin 3\alpha$      &  $       -0.891(29) \times 10^{ -8}$ \\ 
$  6_{ 1, 5}$ &    $\rho_{JJ} $  &   $P^4P_ap_\alpha$                                           &  $       0.8080(23) \times 10^{ -9}$ \\ 
$  6_{ 1, 5}$ &    $\rho_{JK}$  &   $P^2P_a^3p_\alpha$                                         &  $      0.33885(19) \times 10^{ -6}$ \\ 
$  6_{ 1, 5}$ &    $\rho_{KK}$  &   $P_a^5p_\alpha$                                            &  $      0.82693(47) \times 10^{ -4}$ \\ 
$  6_{ 0, 6}$ &    $\Phi_J     $  &   $P^6$                                                      &  $      -0.7082(40) \times 10^{-12}$ \\ 
$  6_{ 0, 6}$ &    $\Phi_{JK}  $  &   $P^4P_a^2$                                                 &  $       0.4015(15) \times 10^{ -9}$ \\ 
$  6_{ 0, 6}$ &    $\Phi_{KJ} $  &   $P^2P_a^4$                                                 &  $      0.84566(56) \times 10^{ -7}$ \\ 
$  6_{ 0, 6}$ &    $\Phi_K$  &   $P_a^6$                                                    &  $     0.141352(69) \times 10^{ -4}$ \\ 
$  6_{ 0, 6}$ &    $2\phi_J $  &   $P^4(P_b^2-P_c^2)$                                         &  $       0.1743(20) \times 10^{-12}$ \\ 
$  6_{ 0, 6}$ &    $2\phi_{JK} $  &   $(1/2)P^2\{P_a^2{,}(P_b^2-P_c^2)\}$                        &  $       0.2724(70) \times 10^{-10}$ \\ 
$  6_{ 0, 6}$ &    $2\phi_K $  &   $(1/2)\{P_a^4{,}(P_b^2-P_c^2)\}$                           &  $        0.124(11) \times 10^{ -9}$ \\ 
$  6_{ 0, 6}$ &    $D_{b2c2bc} $  &   $(1/2)(\{P_b^4{,}P_c^2\}-\{P_b^2{,}P_c^4\})$               &  $       -0.862(14) \times 10^{-12}$ \\ 
$  6_{ 0, 6}$ &    $D_{abJJ} $  &   $(1/2)P^4\{P_a{,}P_b\}$                                    &  $       -0.443(18) \times 10^{-11}$ \\ 
$  6_{ 0, 6}$ &    $D_{abJK} $  &   $(1/2)P^2\{P_a^3{,}P_b\}$                                  &  $       -0.618(72) \times 10^{-10}$ \\ 
$  8_{ 8, 0}$ &    $F_{mmm} $  &   $p_\alpha^8$                                               &  $      0.10917(86) \times 10^{ -8}$ \\ 
$  8_{ 6, 2}$ &    $V_{9J}  $  &   $P^2(1-\cos 9\alpha)$                                      &  $       0.1605(52) \times 10^{ -3}$ \\ 
$  8_{ 6, 2}$ &    $V_{9K}  $  &   $P_a^2(1-\cos 9\alpha)$                                    &  $       0.3053(36) \times 10^{ -1}$ \\ 
$  8_{ 6, 2}$ &    $F_{mmJ} $  &   $P^2p_\alpha^6$                                            &  $        0.484(11) \times 10^{-10}$ \\ 
$  8_{ 6, 2}$ &    $F_{mmK} $  &   $P_a^2p_\alpha^6$                                          &  $      -0.4440(33) \times 10^{ -8}$ \\ 
$  8_{ 6, 2}$ &    $D_{9ac}  $  &   $(1/2)\{P_a{,}P_c\}\sin 9\alpha$                           &  $       0.1404(17) \times 10^{ -1}$ \\ 
$  8_{ 5, 3}$ &    $\rho_{mmJ}  $  &   $P^2P_ap_\alpha^5$                                         &  $        0.928(23) \times 10^{-10}$ \\ 
$  8_{ 5, 3}$ &    $\rho_{mmbc}$  &   $(1/2)\{P_a{,}(P_b^2-P_c^2)\}p_\alpha^5$                   &  $      -0.1085(80) \times 10^{-11}$ \\ 
$  8_{ 5, 3}$ &    $D_{6b2cm}  $  &   $(1/2)\{P_b^2{,}P_c{,}p_\alpha{,}\sin 6\alpha\}$           &  $       -0.977(48) \times 10^{ -7}$ \\ 
$  8_{ 4, 4}$ &    $V_{6JK}  $  &   $P^2P_a^2(1-\cos 6\alpha)$                                 &  $      -0.8483(86) \times 10^{ -6}$ \\ 
$  8_{ 4, 4}$ &    $V_{6bcK} $  &   $(1/2)\{P_a^2{,}(P_b^2-P_c^2)\}(1-\cos 6\alpha)$           &  $      -0.1329(47) \times 10^{ -5}$ \\ 
$  8_{ 4, 4}$ &    $V_{6b2c2} $  &   $(1/2)\{P_b^2{,}P_c^2\}\cos 6\alpha$                       &  $      -0.1068(11) \times 10^{ -7}$ \\ 
$  8_{ 4, 4}$ &    $V_{6abK} $  &   $(1/2)\{P_a^3{,}P_b\}(1-\cos 6\alpha)$                     &  $       -0.462(12) \times 10^{ -6}$ \\ 
$  8_{ 4, 4}$ &    $F_{mJK} $  &   $P^2P_a^2p_\alpha^4$                                       &  $        0.466(13) \times 10^{-10}$ \\ 
$  8_{ 4, 4}$ &    $F_{mKK} $  &   $P_a^4p_\alpha^4$                                          &  $       0.8020(55) \times 10^{ -8}$ \\ 
$  8_{ 4, 4}$ &    $D_{6bcK}  $  &   $(1/2)\{P_a^2{,}P_b{,}P_c\}\sin 6\alpha$                   &  $        0.107(11) \times 10^{ -5}$ \\ 
$  8_{ 4, 4}$ &    $D_{3acmJ}  $  &   $(1/2)P^2\{P_a{,}P_c{,}p_\alpha^2{,}\sin 3\alpha\}$        &  $        0.464(13) \times 10^{ -8}$ \\ 
$  8_{ 4, 4}$ &    $D_{3bcmK}  $  &   $(1/2)\{P_a^2{,}P_b{,}P_c{,}p_\alpha^2{,}\sin 3\alpha\}$   &  $      -0.2284(51) \times 10^{ -8}$ \\ 
$  8_{ 2, 6}$ &    $V_{3JJJ}    $  &   $P^6(1-\cos 3\alpha)$                                      &  $      -0.5682(45) \times 10^{-12}$ \\ 
$  8_{ 2, 6}$ &    $V_{3JJK}    $   &   $P^4P_a^2(1-\cos 3\alpha)$                                 &  $       -0.285(15) \times 10^{-10}$ \\ 
$  8_{ 2, 6}$ &    $V_{3KKK}    $  &   $P_a^6(1-\cos 3\alpha)$                                    &  $       0.3046(67) \times 10^{ -8}$ \\ 
$  8_{ 2, 6}$ &    $V_{3b2c2J}  $  &   $(1/2)P^2\{P_b^2{,}P_c^2\}\cos 3\alpha$                    &  $      -0.3766(36) \times 10^{-11}$ \\ 
$  8_{ 2, 6}$ &    $V_{3b2c2K}  $  &   $(1/2)\{P_a^2{,}P_b^2{,}P_c^2\}\cos 3\alpha$               &  $      -0.3881(93) \times 10^{ -9}$ \\ 
$  8_{ 2, 6}$ &    $V_{3abJJ}   $  &   $(1/2)P^4\{P_a{,}P_b\}(1-\cos 3\alpha)$                    &  $        0.966(22) \times 10^{-11}$ \\ 
$  8_{ 2, 6}$ &    $V_{3abKK}   $  &   $(1/2)\{P_a^5{,}P_b\}(1-\cos 3\alpha)$                     &  $        0.315(11) \times 10^{ -8}$ \\ 
$  8_{ 2, 6}$ &    $F_{JJK}     $  &   $P^4P_a^2p_\alpha^2$                                       &  $        0.741(22) \times 10^{-13}$ \\ 
$  8_{ 2, 6}$ &    $F_{KKK}     $  &   $P_a^6p_\alpha^2$                                          &  $     -0.11704(75) \times 10^{ -7}$ \\ 
$  8_{ 2, 6}$ &    $D_{3acJK}   $  &   $(1/2)P^2\{P_a^3{,}P_c\}\sin 3\alpha$                      &  $      -0.2798(77) \times 10^{ -8}$ \\ 
$  8_{ 2, 6}$ &    $D_{3bcJJ}   $  &   $(1/2)P^4\{P_b{,}P_c\}\sin 3\alpha$                        &  $       0.2201(79) \times 10^{-12}$ \\ 
$  8_{ 2, 6}$ &    $D_{3bcbcJ}  $  &   $(1/2)P^2(\{P_b^3{,}P_c\}-\{P_b{,}P_c^3\})\sin 3\alpha$    &  $       0.1724(14) \times 10^{-11}$ \\ 
$  8_{ 1, 7}$ &    $\rho_{KKK}  $  &   $P_a^7p_\alpha$                                            &  $      -0.9200(56) \times 10^{ -8}$ \\ 
$  8_{ 1, 7}$ &    $\rho_{bcJJ} $  &   $(1/2)P^4\{P_a{,}(P_b^2-P_c^2)\}p_\alpha$                  &  $        0.281(24) \times 10^{-14}$ \\ 
$  8_{ 0, 8}$ &    $L_{KKJ}     $  &   $P^2P_a^6$                                                 &  $      -0.1362(47) \times 10^{-11}$ \\ 
$  8_{ 0, 8}$ &    $L_{K}       $  &   $P_a^8$                                                    &  $      -0.2175(12) \times 10^{ -8}$ \\ 
$ 10_{ 8, 2}$ &    $F_{mmmK}    $  &   $P_a^2p_\alpha^8$                                          &  $       0.1511(67) \times 10^{-12}$ \\ 
$ 10_{ 7, 3}$ &    $\rho_{9bc}  $  &   $(1/2)\{P_a{,}P_b{,}P_c{,}p_\alpha{,}\sin 9\alpha\}$       &  $       -0.805(13) \times 10^{ -5}$ \\ 
$ 10_{ 6, 4}$ &    $F_{mmKK}    $  &   $P_a^4p_\alpha^6$                                          &  $       -0.881(46) \times 10^{-13}$ \\ 
$ 10_{ 6, 4}$ &    $D_{9bcJ}    $  &   $(1/2)P^2\{P_b{,}P_c\}\sin 9\alpha$                        &  $       0.6674(82) \times 10^{ -8}$ \\ 
$ 10_{ 4, 6}$ &    $V_{6JJK}    $  &   $P^4P_a^2(1-\cos 6\alpha)$                                 &  $       -0.307(16) \times 10^{-10}$ \\ 
$ 10_{ 4, 6}$ &    $V_{6KKK}    $  &   $P_a^6(1-\cos 6\alpha)$                                    &  $      -0.5642(92) \times 10^{ -8}$ \\ 
$ 10_{ 4, 6}$ &    $F_{mabJK}   $  &   $(1/2)P^2\{P_a^3{,}P_b\}p_\alpha^4$                        &  $        0.547(27) \times 10^{-14}$ \\ 
$ 10_{ 2, 8}$ &    $V_{3JJJK}   $  &   $P^6P_a^2(1-\cos 3\alpha)$                                 &  $        0.756(59) \times 10^{-15}$ \\ 
$ 10_{ 2, 8}$ &    $V_{3KKKK}   $  &   $P_a^8(1-\cos 3\alpha)$                                    &  $       -0.284(22) \times 10^{-11}$ \\ 
$ 10_{ 2, 8}$ &    $V_{3b2c2JJ} $  &   $(1/2)P^4\{P_b^2{,}P_c^2\}\cos 3\alpha$                    &  $       -0.146(12) \times 10^{-15}$ \\ 
$ 12_{ 8, 4}$ &    $V_{12JK}    $  &   $P^2P_a^2(1-\cos 12\alpha)$                                &  $       0.1899(28) \times 10^{ -4}$ \\ 
              &    $\chi_{aa}$  &                                                              &  $     -0.27527(18) \times 10^{ -3}$ \\ 
              &    $\chi_{bb}$  &                                                              &  $      -0.9514(37) \times 10^{ -4}$ \\ 
              &    2$\chi_{ab}$ &                                                              &  $       -0.261(29) \times 10^{ -3}$ \\ 

\hline
\hline

\end{longtable}

\tablefoot{$^{a}$ \textit{n=t+r}, where \textit{n} is the total order of the operator, \textit{t} is the order 
  of the torsional part, and \textit{r} is the order of the rotational part, respectively. The ordering scheme 
  of \citet{Nakagawa:1987} is used. $^{b}$ The parameter nomenclature is based on the subscript 
  procedure of \citet{CH3OH_rot_2008}. $^{c}$ $\lbrace A,B,C,D,E \rbrace = ABCDE+EDCBA$.  $\lbrace A,B,C,D \rbrace = ABCD+DCBA$. $\lbrace A,B,C \rbrace = ABC+CBA$. $\lbrace A,B \rbrace = AB+BA$. 
  The product of the operator in the third column of a given row and the parameter in the second column 
  of that row gives the term actually used in the torsion-rotation Hamiltonian of the program, except for 
  \textit{F}, $\rho$, and \textit{$A_{\rm RAM}$}, which occur in the Hamiltonian in the form 
  $F(p_\alpha + \rho P_a)^2 + A_{\rm RAM}P_a^2$.  $^{d}$ Values of the parameters in units of cm$^{-1}$, except for $\rho$, 
  which is unitless. $^{e}$ Statistical uncertainties are given in parentheses as one standard uncertainty 
  in units of the last digits.}


\vspace{\fill}

\section{Additional figures from the ReMoCA survey}
\label{a:remoca}

Figures~\ref{f:remoca_ch3oh_13c_ve0_n2b}--\ref{f:remoca_ch3oh_18o_ve1_n2b}
show the transitions of $^{13}$CH$_3$OH $\varv = 0$, 
$^{13}$CH$_3$OH $\varv_{\rm t} = 1$, CH$_3$$^{18}$OH $\varv = 0$, and
CH$_3$$^{18}$OH $\varv_{\rm t} = 1$, respectively, that are covered by the 
ReMoCA survey and significantly contribute to the signal detected toward 
Sgr~B2(N2b). Transitions that are too heavily blended with much stronger 
emission (or absorption) from other molecules and therefore cannot contribute 
to the identification of these isotopologs are not shown in these figures.
Figures~\ref{f:popdiag_ch3oh_13c_n2b}--\ref{f:popdiag_ch3oh_17o_n2b} show
the population diagrams obtained for $^{13}$CH$_3$OH, CH$_3$$^{18}$OH, and
CH$_3$$^{17}$OH, respectively.

\begin{figure*}
\centerline{\resizebox{0.85\hsize}{!}{\includegraphics[angle=0]{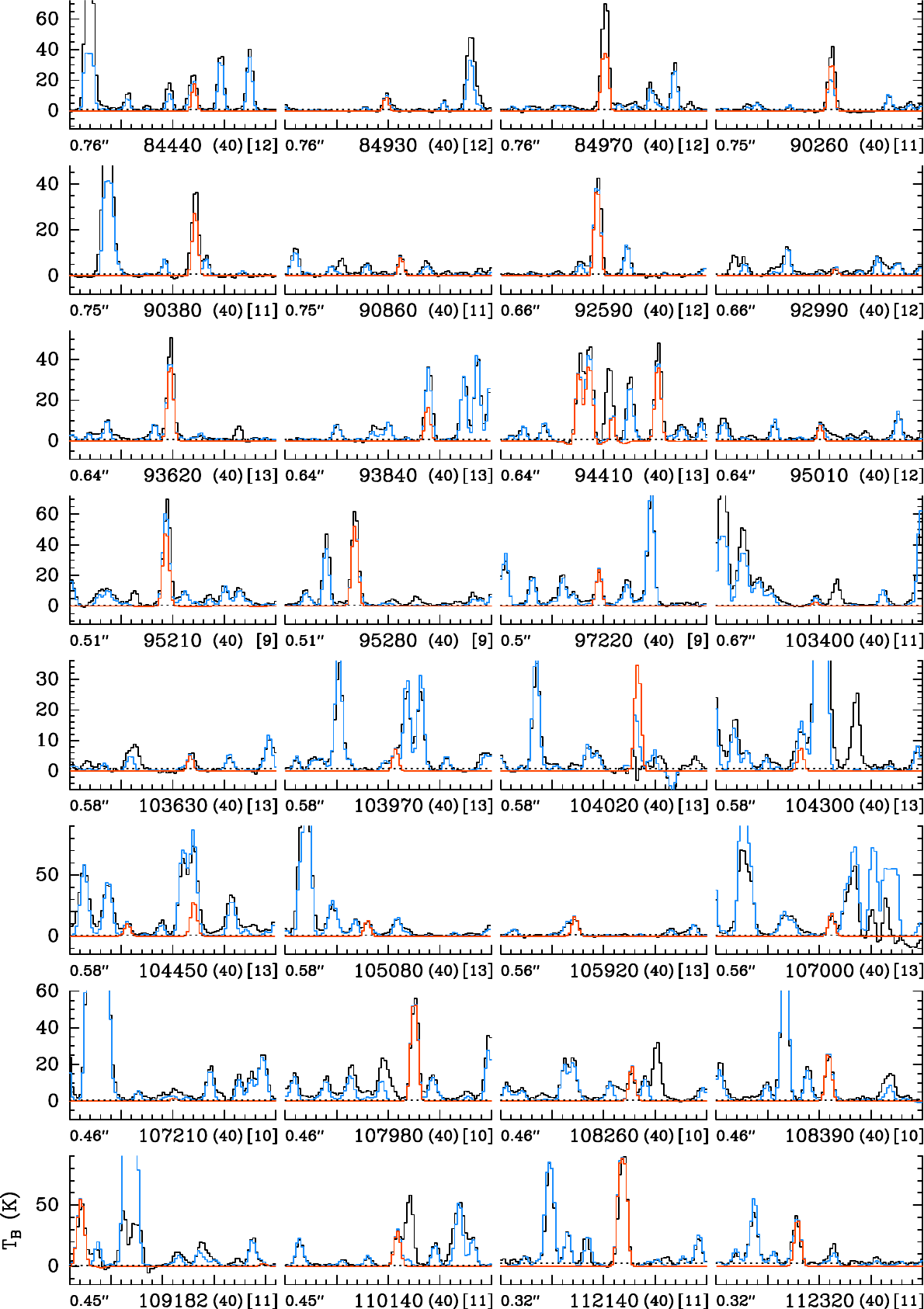}}}
\caption{Selection of rotational transitions of $^{13}$CH$_3$OH $\varv = 0$ 
covered by the ReMoCA survey. The LTE 
synthetic spectrum of $^{13}$CH$_3$OH $\varv = 0$  is displayed in red and 
overlaid on the observed spectrum of Sgr~B2(N2b) shown in black. The blue 
synthetic spectrum contains the contributions of all molecules identified in 
our survey so far, including the contribution of the species shown in red. The 
values written below each panel correspond from left to right to the half-power 
beam width, the central frequency in MHz, the width in MHz of each panel in 
parentheses, and the continuum level in K of the baseline-subtracted spectra 
in brackets. The y-axis is labeled in brightness temperature units (K). The 
dotted line indicates the $3\sigma$ noise level.}
\label{f:remoca_ch3oh_13c_ve0_n2b}
\end{figure*}

\begin{figure*}
\addtocounter{figure}{-1}
\centerline{\resizebox{0.46\hsize}{!}{\includegraphics[angle=0]{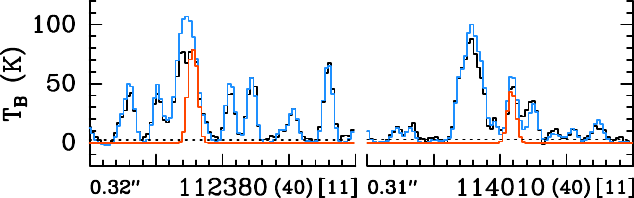}}}
\caption{continued.}
\end{figure*}

\begin{figure*}
\centerline{\resizebox{0.85\hsize}{!}{\includegraphics[angle=0]{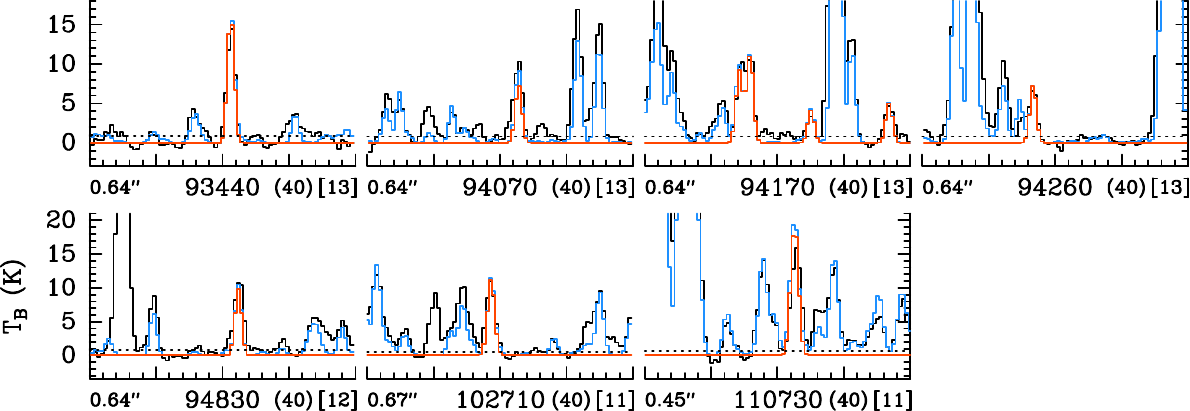}}}
\caption{Same as Fig.~\ref{f:remoca_ch3oh_13c_ve0_n2b}, but for
$^{13}$CH$_3$OH $\varv_{\rm t} = 1$.}
\label{f:remoca_ch3oh_13c_ve1_n2b}
\end{figure*}

\begin{figure*}
\centerline{\resizebox{0.85\hsize}{!}{\includegraphics[angle=0]{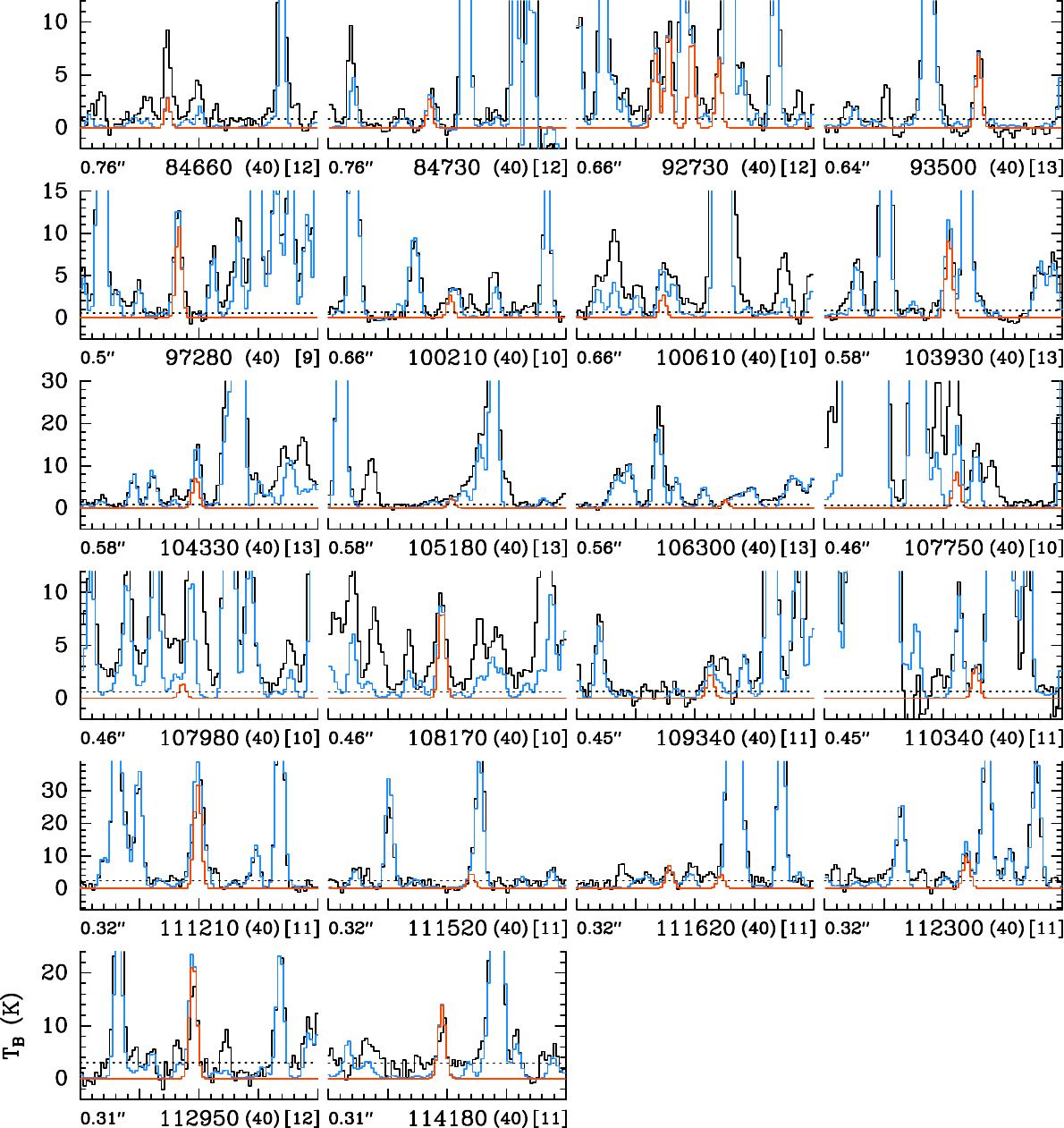}}}
\caption{Same as Fig.~\ref{f:remoca_ch3oh_13c_ve0_n2b}, but for
CH$_3$$^{18}$OH $\varv = 0$.}
\label{f:remoca_ch3oh_18o_ve0_n2b}
\end{figure*}

\begin{figure*}
\centerline{\resizebox{0.85\hsize}{!}{\includegraphics[angle=0]{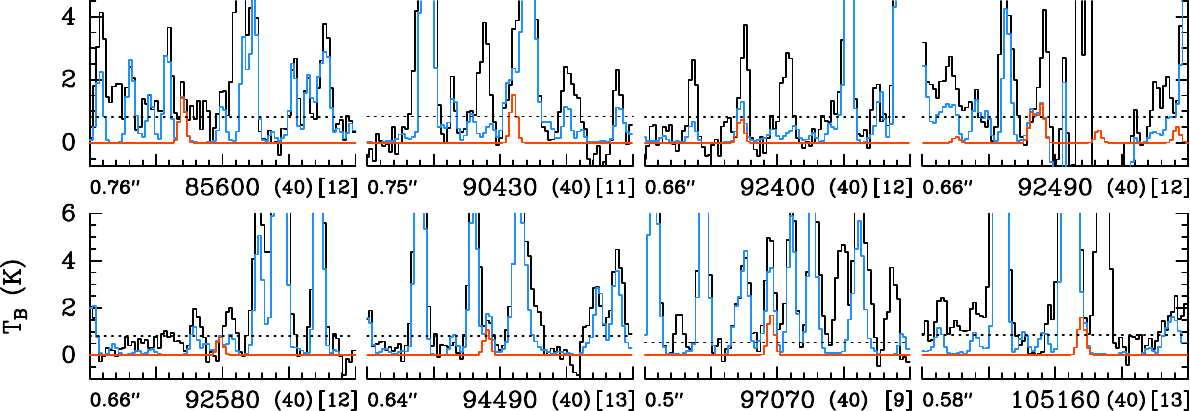}}}
\caption{Same as Fig.~\ref{f:remoca_ch3oh_13c_ve0_n2b}, but for
CH$_3$$^{18}$OH $\varv_{\rm t} = 1$.}
\label{f:remoca_ch3oh_18o_ve1_n2b}
\end{figure*}

\begin{figure}[!h]
\centerline{\resizebox{0.5\hsize}{!}{\includegraphics[angle=0]{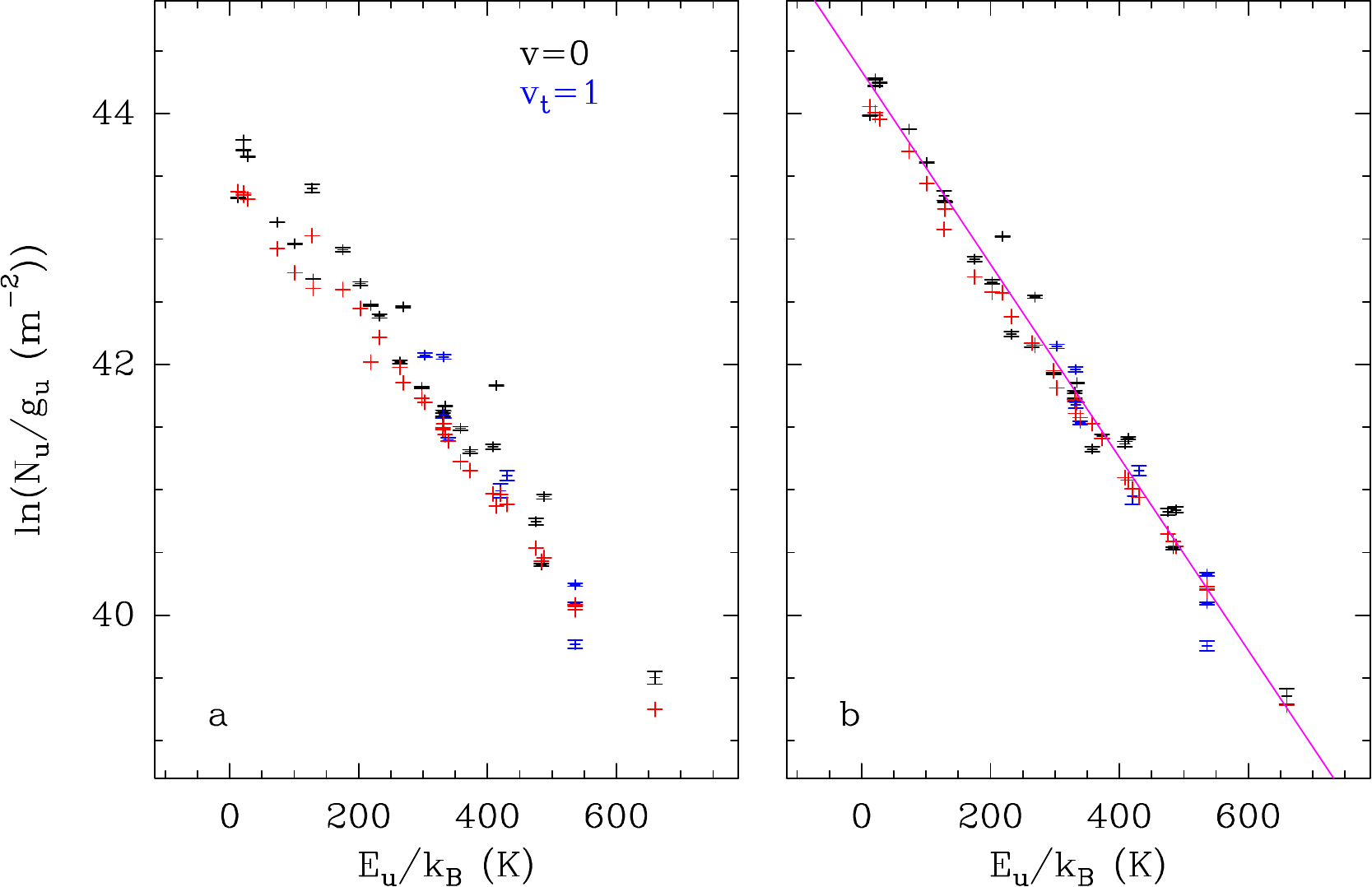}}}
\caption{Population diagram of $^{13}$CH$_3$OH toward Sgr~B2(N2b). The
observed data points are shown in black while the synthetic
populations are shown in red. No correction is applied in panel {\bf a}.
In panel {\bf b}, the optical depth correction has been applied to both the
observed and synthetic populations and the contamination by all other
species included in the full model has been subtracted from the observed
data points. The purple line is a linear fit to the observed populations (in
linear-logarithmic space).
}
\label{f:popdiag_ch3oh_13c_n2b}
\end{figure}

\begin{figure}[!h]
\centerline{\resizebox{0.5\hsize}{!}{\includegraphics[angle=0]{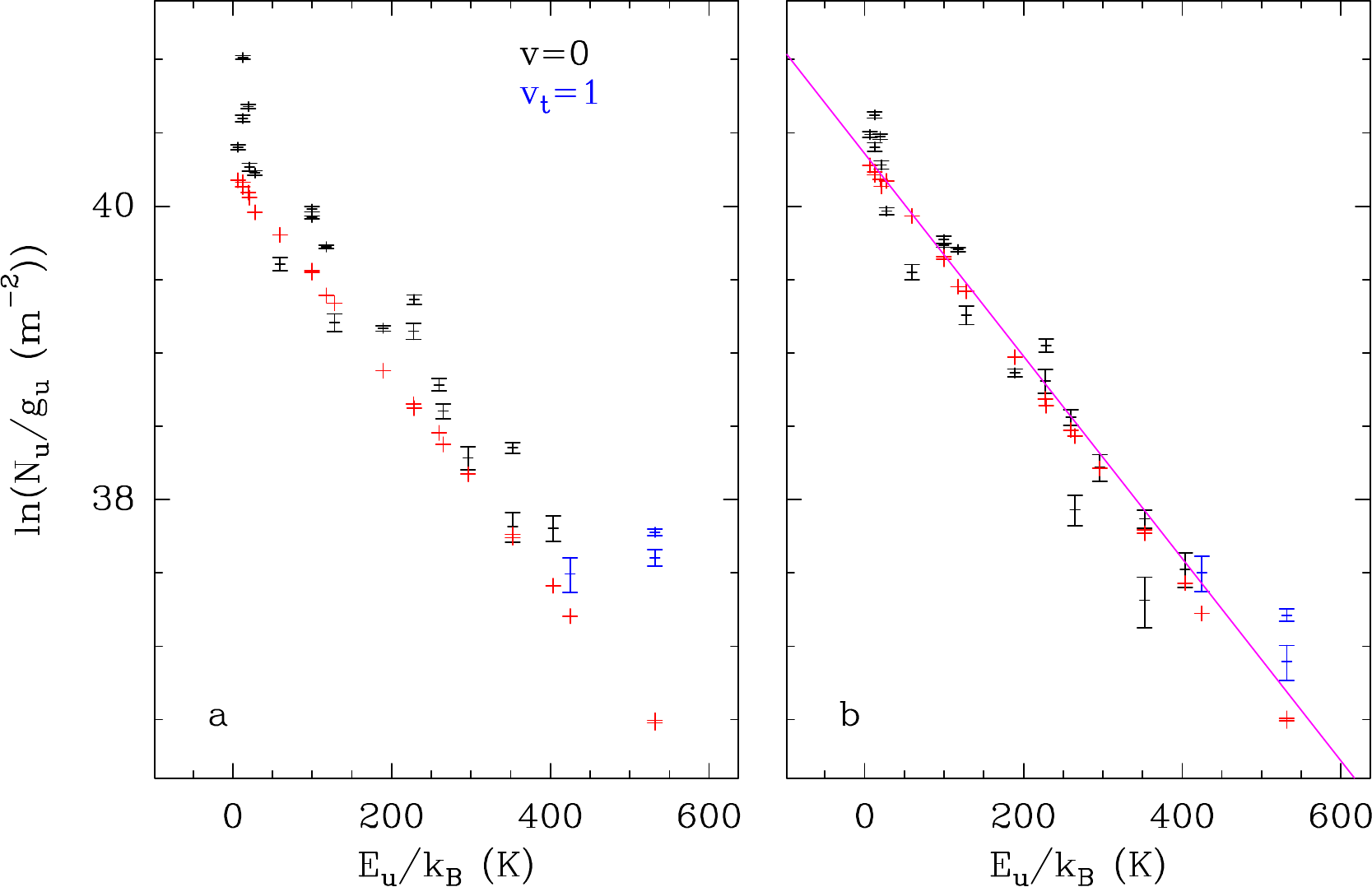}}}
\caption{Same as Fig.~\ref{f:popdiag_ch3oh_13c_n2b}, but for CH$_3$$^{18}$OH.
}
\label{f:popdiag_ch3oh_18o_n2b}
\end{figure}

\begin{figure}[!h]
\centerline{\resizebox{0.5\hsize}{!}{\includegraphics[angle=0]{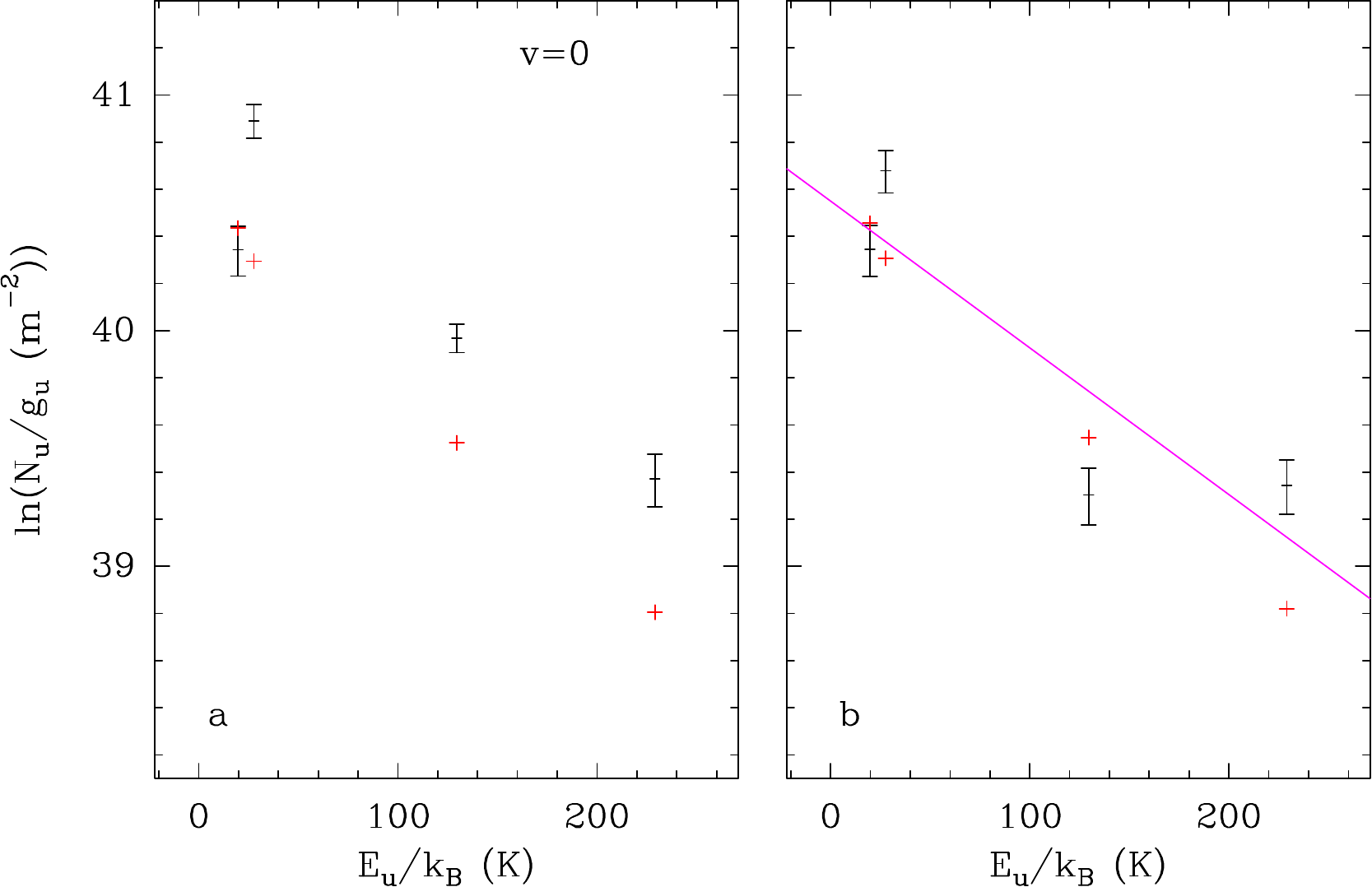}}}
\caption{Same as Fig.~\ref{f:popdiag_ch3oh_13c_n2b}, but for CH$_3$$^{17}$OH.
}
\label{f:popdiag_ch3oh_17o_n2b}
\end{figure}

\end{appendix}
\end{document}